\documentclass[traditabstract]{aa} 
\usepackage{graphicx}
\usepackage{txfonts}
\usepackage{textcomp}
\usepackage{url}

\usepackage{natbib}
\bibpunct{(}{)}{;}{a}{}{,} 

\usepackage{txfonts}

\usepackage{rotating}
\usepackage{amssymb}
\begin{document}
  \title{Carbon and nitrogen abundances of stellar populations in the globular cluster M~2\thanks{Based on observations 
made with the Italian Telescopio Nazionale Galileo (TNG) operated on the island of La Palma by the 
Fundaci\'{o}n Galileo Galilei of the INAF (Istituto Nazionale di Astrofisica) at the Spanish Observatorio del 
Roque de los Muchachos of the Instituto de Astrofisica de Canarias (PROGRAM ID: A22TAC\_20).}\fnmsep\thanks{
Table~\ref{PHOTOSAMPLE} is only available in electronic form
at the CDS via anonymous ftp to cdsarc.u-strasbg.fr (130.79.128.5)
or via http://cdsweb.u-strasbg.fr/cgi-bin/qcat?J/A+A/}}


   \author{C.~Lardo
          \inst{1}, E.~Pancino\inst{2}, A.~Mucciarelli\inst{1}, and A.~P.~Milone\inst{3,4}
          }
  
  \authorrunning{Lardo et al.}

   \institute{Department of Astronomy, University of Bologna, Via Ranzani 1,
 40127 Bologna, Italy;~\email{carmela.lardo2@unibo.it}\\
        \and INAF-Osservatorio Astronomico di Bologna, Via Ranzani 1, 40127 Bologna, Italy\\
        \and Instituto de Astrof\'{\i}sica de Canarias, E-38200 La Laguna, Tenerife, Canary Islands, Spain\\
	\and Department of Astrophysics, University of La Laguna, E-38200 La Laguna, Tenerife, Canary Islands, Spain}

   \date{Received/Accepted}

 
 \abstract {
We present CH and CN index analysis and C and N abundance calculations based on the low-resolution blue spectra of 
red giant branch (RGB) stars in the Galactic globular cluster NGC 7089 (M~2). Our main goal is to investigate the 
C-N anticorrelation for this intermediate metallicity cluster.
The data were collected with DOLORES, the multiobject, low-resolution facility at the Telescopio Nazionale Galileo.
We first looked for CH and CN band strength variations and bimodalities in a sample of RGB stars with
17.5 $\leq V \leq$ 14.5. Thus we derived C and N abundances under LTE assumption by comparing observed spectra
with synthetic models from the spectral features at 4300\AA~(G-band) and at $\sim$3883\AA~(CN).
Spectroscopic data were coupled with $UV$ photometry obtained during the spectroscopic run.
We found a considerable star-to-star variation in both A(C) and A(N) at all luminosities for our sample of 35 targets. 
These abundances appear to be anticorrelated, with a hint of bimodality in the C content for stars with 
luminosities below the RBG bump (V$\sim$15.7), while the range of variations in N abundances is very large and spans almost 
$\sim$ 2 dex.
We find additional C depletion as the stars evolve off the RGB bump, in fairly good agreement with theoretical predictions
for metal-poor stars in the course of normal stellar evolution.
We isolated two groups with N-rich and N-poor stars and found that N abundance variations correlate with the 
$(U-V)$ color in the DOLORES color-magnitude diagram (CMD). The $V, (U-V)$ CMD for this cluster shows an additional RGB sequence,
located at the red of the main RGB and amounting to a small fraction of the total giant population.
We identified two CH stars detected in previous studies in our $U, V$ images.
These stars, which are both cluster members, fall on this redder sequence, 
suggesting that the anomalous RGB should have a peculiar chemical pattern. 
Unfortunately, no additional spectra were obtained for stars in this previously unknown RGB branch.}
{}

\maketitle
\section{Introduction}\label{introduzione}
In the past few years, a large collection of spectroscopic and photometric data have conclusively determined that 
globular clusters (GCs) can no longer be considered systems made up of a simple monometallic population.
GCs are largely homogeneous with regard to iron and $n$-capture elements 
(with the outstanding exceptions of $\omega$ Cen, M~22, M~54, Ter~5, and NGC~1851;
see  \citealp{jhonson10}; \citealp{marino12}; \citealp{carretta54}; \citealp{origlia11}; \citealp{yong08}), 
while they show a significant spread in the abundance of lighter
elements involved in proton-capture processes, with strong anticorrelations between the abundances of C and N, Na and O, or
Mg and Al, as well as bimodal distribution of CH and CN band strength
\citep[][among others]{kraft94,cohen02,ramirez03,cohen05,carretta2009,mart09,kayser08,pancino10}.
These variations are not observed in field counterparts of the same metallicity\footnote{Using moderate-resolution 
spectra of 561 giants with typical halo metallicities, \citet{martellgrebel} find that 3\% of the sample shows
the CN-CH bandstrength typical of GC stars. They argue that  
these stars are genuine second-generation GC stars that have 
been lost to the halo field.} \citep[but they show signs of
dredged-up CNO processing,][]{gratton00} or in open clusters \citep{desilva09,pancinoopen10}.

This peculiar chemical pattern appears to be ubiquitous for all GCs that have been studied properly.
Originally, the first detection of unusual abundances came from the bright red giant branch (RGB) stars:
spectroscopic investigations of the CH and CN absorption features
often revealed a bimodality in the CN band strength that is accompanied by a broader distribution of CH
\citep[][and references therein]{kayser08,pancino10}.
Here we want to concentrate on the carbon and nitrogen abundance variations for RGB stars.
In spite of the large choice of literature on this topic, many questions still remain open in the understanding of how 
the observed chemical variations of C and N formed.

Following the classical prediction, during the H-burning phase 
via the CNO cycle, N is enriched at the cost of C and O. 
When a star evolves off the main sequence, the convective envelope starts to move inward, 
{\em dredging up} material that has been processed through
partial hydrogen burning by the CNO cycle and pp chains.
Canonically, light-element abundances should be untouched by subsequent evolution along the RGB,
but the observational evidence has shown that both various light-element abundances (particularly [C/Fe] and log $\epsilon$(Li))
and isotopic ratios (\element[][12]{C}/\element[][13]{C}) vary
as the stars evolve along the RGB, and this cannot be accounted 
by a single first dredge-up alone.

Some further nonconvective {\em deep mixing} should take place in the advanced phases 
of RGB evolution: after the end of the dredge-up phase is reached, the star's convective envelope
begins to move outward, leaving behind a sharp discontinuity in mean molecular weight (the $\mu$-barrier) at the point of
deepest inward progress \citep{iben68}.
The corresponding change in molecular weight can potentially hinder further mixing.
However, during the evolution along the RGB, the hydrogen-burning shell advances outward and eventually encounters 
the $\mu$-barrier. The influx of fresh hydrogen-rich material to the hydrogen-burning shell causes a temporary slowdown 
of the star's evolution, which manifests itself in a bump in the differential luminosity function (LF) of the cluster.
Thereafter, since the molecular gradient is effectively canceled out, some further mixing episodes are allowed. 
Briefly, possible sources of extramixing could be rotation-induced mixing \citep{charbonnel95}
or thermohaline mixing associated with the reaction 3He(3He,2p)4He \citep{angelou12}.
Extramixing is a universal mechanism that occurs in $\geq$ 96\% of 
these RGB bump stars \citep{charbonnel98} in the field, in open and globular clusters
and also in stars in external galaxies.
As a consequence, normal stellar evolution contributes to the C-N anticorrelation observed among bright RGB stars.

However, mixing cannot be the only driving mechanism of these abundance variations, since
large star-to-star light-element variations are also observed among RGB stars at the same evolutionary stage, and
they are also observed in unevolved stars \citep{cannon03,ramirez03,cohen05}, indicating the occurrence of high-temperature
hydrogen-burning processes (CNO, Ne-Na, Mg-Al cycles) that cannot occur in low-mass GC stars \citep{gratton01}.
Therefore, the peculiar chemistry observed should also have an {\em external} origin.
The so-called {\em multiple populations} scenario foresees the occurrence of at least two
episodes of star formation: CN-weak stars being the first stars that formed,
while CN-strong  stars formed some tens/hundreds of Myr later from the enriched 
ashes of the first generation \citep{dercole08}. 
Up to now, we lack a complete understanding of the current 
mechanism that drives the observational facts, although theoretical nucleosynthesis models show that 
the the observed chemical pattern can be provided either by 
intermediate-mass (M $>$3-5 M\sun) asymptotic giant branch (AGB) stars \citep{ventura08} 
or fast-rotating massive stars \citep[FRMS,][]{decressin07}.
The detection of a bimodal distribution of CN band strength below the bump in the LF, 
in moderate-metallicity clusters ([Fe/H] $\geq$ --1.6), provides strong evidence for this 
scenario \citep{briley91,cannon98,kayser08,pancino10,smolinski11}.
Furthermore, mixing must be operating in each generation \citep{suntzeff91, 
denissenkov98} as stars evolve along the RGB.
In addition to these two main scenarios, there are several different mechanisms that can potentially produce 
chemical inhomogeneities in GCs. 
We refer the reader to the comprehensive review by \citet{gratton12} for a discussion.

In recent years, thanks to the large capabilities of the Hubble space telescope (HST), 
accurate photometry conclusively demonstrated that a growing number of GCs host at least two stellar generations of stars.
Indeed, star-to-star variations in light- and alpha-element abundances, age, and metallicity can 
determine multimodal or broad sequences in the CMD observed within 
some galactic or extragalactic GCs 
(e.g.,~\citealp{pancino00,bedin04,sollima07,piotto07,marino08,milone10,lardo11}).
Even if it is not yet clear how photometric complexity be mapped for the variations in age/or chemical abundance, there are several 
tempting potential connections between the photometric multiplicity and light-element abundance inhomogeneities (e.g., see 
\citealp{marino09} for M~22, and \citealp{yong08}, \citealp{yong09}, \citealp{gratton12} 
and \citealp{lardo12} for the case of NGC~1851).  

With this paper we investigate the behavior of carbon and nitrogen along the RGB of M~2.
This is an intermediate-metallicity ([Fe/H] =--1.65; \citealp{harris96}, 2010 edition) cluster, 
which is located 11.5 kpc from the galactic center, is relatively rich, and lies in a sparse field.
This cluster was found by \citet{smith90} to have a bimodal CN distribution, 
with the majority of red giants found to be CN-strong stars. 
Earlier works have already revealed a large number of stars with strong 
$\lambda$3883 CN bands \citep{mcclure81, canterna82,smith90}.
Furthermore, this cluster is found to contain CH stars \citep{smith90,zinn81}.
In a recent paper with a large sample of stars in all evolutionary phases, \citet{smolinski11} detected
signs of enhanced N enrichment well before the point of first dredge-up, besides the usual CN variations on the RGB.
On the photometric front, M~2 $g, (u-g)$ CMD from Sloan digital sky survey (SDSS) photometry \citep[see Fig.~1 of][]{lardo11}
shows evidence of a spread in light-element abundances, which comes from the significant spread along the RGB, which
would not be detected in $(g-r)$, and which is incompatible with measurements errors alone or with 
differential reddening effects.
The broadening in the $U, (U-V)$ CMD (and/or usual visual colors) may be a different way
to search for multiple stellar populations. 
In these respects, \citet{marino08} find that the Na (or O)
distribution is bimodal, with a rich sample of more than 100 giants in M~4, and this bimodal distribution is 
correlated with a bimodal distribution in CN strength, too (Na-rich stars are also CN strong), which is 
associated to a dichotomy in the color of RGB stars 
in the $U, (U-B)$ CMD. 
Prompted by these considerations, we thought that $U$-based photometry can be used, when coupled with C and N abundances 
from analyzing low-resolution spectra, for efficient tagging of multiple stellar populations in M~2.

This article is structured as follows.
We describe the sample in Sect.~\ref{preimaging}.
We outline our measurements of the CN and CH indices and their interpretation in Sect.~\ref{misura_indici}.
We derive C and N abundances from spectral synthesis in Sect.~\ref{analisi_abbondanze} and discuss the result in 
Sect.~\ref{risultati}. 
We discuss and analyze the split of the RGB discovered in the $V,U-V$ CMD of M~2 in Sect.~\ref{rgb_anomalo}.
Finally we present a summary of our results and draw conclusions in Sect.~\ref{conclusioni}.

\section{Observational material}\label{preimaging}
We selected M~2 spectroscopic targets from the \citet{an08} publicly available photometry.
\citet{an08} reanalyzed SDSS images of the GCs (and open clusters) included in the survey using 
the DAOPHOT/ALLFRAME suite of programs \citep{stetson87, stetson94}. 
In our previous work \citep{lardo11}, we used \citet{an08} photometry to search
for anomalous spread in near UV color ({\it u-g}) along the RGB of nine
Galactic GCs and study the radial profile of the first and second generation stars.
We refer the reader these two papers for a detailed description of the photometric database employed to select 
spectroscopic targets.
The initial sample of candidate stars consisted of those located more than 
1\arcmin\ from the center of M~2
(to facilitate sky subtraction) with 14.5 $< V <$ 17.5~mag. 
Spectroscopic targets were hence chosen as the most isolated stars (no neighbors within 2\arcsec) 
as close as possible to the main locus of the RGB sequence in the $g, (u-g)$ and $g,(g-r)$ diagrams to reduce 
the incidence of blended images\footnote{Unfortunately, by using these selection criteria, we accidentally excluded
stars belonging to previously unknown additional RGB sequence (see Sect.~\ref{fotometria}) from our spectroscopic sample.}.

\subsection {Photometry}\label{fotometria}
In addition, we also obtained images of the cluster in the standard Johnson 
$U$ and $V$ filters for a total of 540 s shifted in three single 
exposures in each filter with the DOLORES camera.
DOLORES (Device Optimized for the LOw RESolution) is a low-resolution spectrograph 
and camera permanently installed at Telescopio Nazionale Galileo (TNG) located in La Palma,
Canary Islands (Spain).
The choice of pass-bands is due to the ability of separating photometric sequences 
at different evolutionary stages along the CMD (as discussed in Sects.~\ref{introduzione} and ~\ref{rgb_anomalo}).
The DOLORES camera offers a field of view (FoV) 
of  8.6\arcmin $\times$ 8.6\arcmin~with a 0.252 arcsec/pix scale.
The raw frames were processed (bias-subtracted and flat-fielded) using the
standard tasks in IRAF \footnote{IRAF is distributed by the National Optical Astronomy Observatory,
which is operated by the Association of Universities for Research in 
Astronomy, Inc., under cooperative agreement with the National
Science Foundation.}.
Point spread function (PSF) fitting photometry was thus carried out with the 
DAOPHOT II and ALLSTAR packages \citep{stetson87,stetson94} using a constant model PSF.
The photometric calibration was done using stars in common with Stetson Photometric standard field \citep{stetson00}\footnote{available at 
{\tiny\tt http://www3.cadc-ccda.hia-iha.nrc-cnrc.gc.ca/\\community/STETSON/standards/}}.
Stars within 1\arcmin~and outside of 4\arcmin~from the cluster center are excluded from the CMD
to reduce blending effects and the field star contamination, respectively.
The rms in magnitude and the chi and sharp parameters are powerful indicators of the photometric 
quality\footnote{On all stars we imposed the selection limits of CHI $<$ 2.0 and -1 $<$ SHARP $<$ 1 on DAOPHOT II photometric parameters. 
The first of these parameters, CHI, is the
ratio of the observed pixel-to-pixel scatter in the fitting residuals to the expected scatter, based on the
values of readout noise and the photons per ADU specified in the DAOPHOT options file, while SHARP is a zeroth-order estimate
of the square of the quantity $SHARP^{2} \sim \sigma^{2}(observed) - \sigma^{2}(point-spread function)$; 
see the DAOPHOTII manual at \url{http://www.astro.wisc.edu/sirtf/daophot2.pdf}.}.
To select a sample of well-measured stars we followed the procedure given in \citet{lardo12}, Sect.~5.1.
The catalog of the selected sample is presented in Table~\ref{PHOTOSAMPLE}.
\begin{table}
\caption{Photometry of M~2: selected sample.}             
\label{PHOTOSAMPLE}      
\centering          
\begin{tabular}{l c c c c c c }
\hline \hline  \\
ID      &     RA      &   Dec         &  {\it V}    & $\epsilon V $ & {\it U--V}      &  $\epsilon U$  \\
	& (deg)       &    (deg)     &    (mag)    &   (mag)       &   (mag)         & (mag)           \\
\hline
    164 & 323.3657998 &  -0.8890510 &   20.215  &   0.026  &   0.213&     0.076 \\
    196 & 323.3800929 &  -0.8885692 &   19.943  &   0.019  &   0.311&     0.072 \\
    234 & 323.3661658 &  -0.8882047 &   20.342  &   0.018  &   0.393&     0.084 \\
    286 & 323.3795809 &  -0.8875061 &   19.015  &   0.012  &   0.220&     0.028 \\
    303 & 323.3872881 &  -0.8872826 &   18.445  &   0.010  &   0.632&     0.023 \\
    313 & 323.3674707 &  -0.8870783 &   19.661  &   0.024  &   0.202&     0.050 \\
    330 & 323.3908254 &  -0.8869171 &   19.613  &   0.019  &   0.267&     0.041 \\
    336 & 323.3657205 &  -0.8868386 &   19.325  &   0.012  &   0.305&     0.037 \\
\hline
\hline                  
\end{tabular}
\tablefoot{A portion of the table is shown for guidance about its content, the complete table is available in electronic 
format through the CDS service. }
\end{table} 
\begin{figure}
  \centering
  \resizebox{\hsize}{!}{\includegraphics{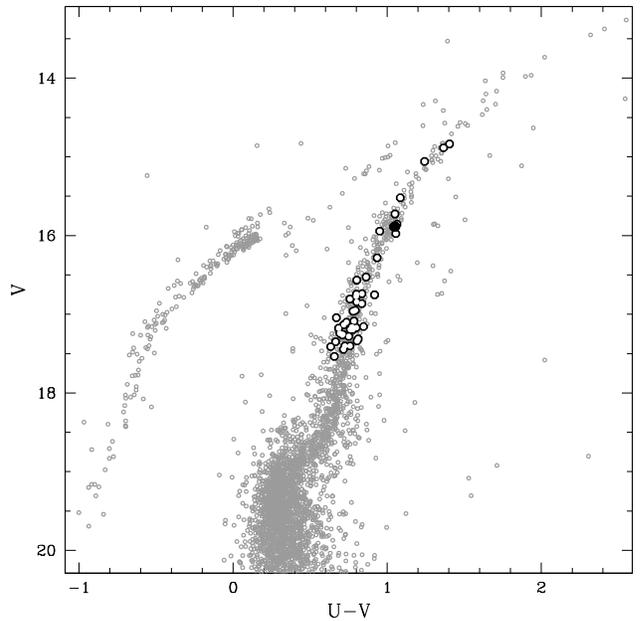}}
\caption{$V, (U-V)$ CMD for M~2 from DOLORES images. White dots mark spectroscopic targets, the black dot 
shows the probable field star (see Sect.~\ref{spettroscopia}).}.
        \label{M2-DOLORES}
   \end{figure}
The resulting calibrated (and corrected for differential reddening) {\it (V, U-V)} CMD for M~2 and the spectroscopic 
target stars are shown in Fig.~\ref{M2-DOLORES}.

\subsection{Spectroscopic observations and reduction}\label{spettroscopia}
Stellar spectra were obtained with DOLORES
which allows for multi-slit spectroscopy.
We defined three slit masks using the stand-alone version of the Interactive Mask Design Interface,
provided by the DOLORES staff at the telescope\footnote{see for reference {\tiny\tt http://web.oapd.inaf.it/mos/}}. 
The positions of the program stars were determined using M~2 catalogs by \citet{an08},  
as discussed in Sect.~\ref{preimaging}.
The slit width on the masks was fixed to
1.1\arcsec, and the slit length was chosen to be at least 8\arcsec to allow
for local sky subtraction. Typically, we succeeded to fit $\simeq$16 slits onto
one mask (for a total of 48 target stars).
Because the goal of our spectroscopic observations was to measure
the strengths of the 3880 and 4300 \AA~CN and CH  absorption
bands, we used the LRB grism with a dispersion of 2.52\AA/pix.
In combination with the chosen slit width this results in the spectral 
resolution of R(@3880\AA)=353 and R(@4305\AA)=391 in the wavelength region of interest.
The grism's spectral region covers the nominal wavelength range between $3000-8430$\AA, while the actual spectral
coverage depends on the location of the slit on the mask with respect to the dispersion direction.
To reach a high S/N, each mask configuration was observed three times with exposure
durations of 1800 s each, leading to a total exposure time of
1.5 hr per mask and a typical S/N of $\simeq$ 20-30 in the CN region.
Additionally, bias, flat field, and wavelength
calibration observations were obtained in the afternoon.

For the data prereduction, we used the standard procedure for overscan correction
and bias-subtraction with the routines available in the \textit{noao.imred.ccdred} package in
IRAF. First, we stacked the flat fields for each night
and mask.  Because each slit mask was observed
three times and the alignment of the frames was quite good, we
co-added the three frames for a given slit mask with cosmic-ray
rejection enabled, providing resulting frames that were almost free
of cosmic rays. For the following analysis we extracted the area
around each slit with the optimal extraction and treated the resulting spectra as single-slit 
observations. The wavelength-calibration images and
flat fields were treated in the same manner. 
TNG spectroscopic flats show a severe internal reflection problem in the blue regions of the spectra 
that could in principle heavily affect 
our further measurements. To minimize this effect, we fit the 2D large-scale
structures in the normalized spectroscopic flat field by smoothing and 
dividing the original flat by the fit, keeping the small pixel-to-pixel variations, which are the ones we
intend to correct for  with flat fielding. 
The object spectra and arc
images were flat-field-calibrated with these corrected
flat fields. Standard IRAF routines were used to wavelength-calibrate,
sky-subtract, and extract the stellar spectra. The
wavelength solution from the HeNeHg arcs was fitted by a
first-order spline. The typical rms of the wavelength calibration
is on the order of 0.3 \AA , which is largely expected at the
given spectral resolution. The residual uncertainties in the wavelength calibration are then removed using the 
position of strong emission lines (in particular OI at 5577.7 \AA~and NaD at 5895 \AA).

The shape of the final spectra is affected by the dependence of the instrumental
response on the wavelength. Given the quite low instrument response in the blue part of the spectrum
and the presence of many absorptions in this region, we avoided any attempt to remove this effect 
through flux calibration or continuum normalization \citep[see][and references therein]{pancino10}.

To derive the membership of candidate RGB stars, we first performed a cross-correlation of 
the object spectrum with the highest S/N star on each MOS mask as a template with the IRAF {\em fxcor} routine.
The template $V_{r}$ were computed using the laboratory positions of the most prominent spectral features
(e.g., $H_{\alpha}$, $H_{\beta}$, $H_{\gamma}$, $H_{\delta}$, and Ca H+K, among others), yielding a mean radial velocity of
-13 $\pm$ 30 km/s for the entire sample. This value, given the low resolution of our spectra, agrees quite well with the value 
tabulated (--5.0 km/s) in the \citealp{harris96} (2010 edition) catalog.
Then, we rejected individual stars with values deviating by more than 3$\sigma$ from this average velocity, 
deeming them to be probable field stars. Only one star (ID:10427, see Fig.~\ref{M2-DOLORES}) 
was rejected based on its radial velocity.
In a final step, we examined each spectrum individually and rejected spectra with defects (like spikes or holes)
in the measurement windows. 

\section{CH and CN band strengths}\label{misura_indici}
A set of indices quantifying the strengths of the UV CN band, the G band 
of CH and the $Ca_{II}$ H and K lines were measured for the spectra.
To be consistent with our previous work, we adopted the indices as defined by \citet{harbeck03} and \citet{pancino10}:
$$S(3839)= -2.5 \log \frac{F_{3861-3884}}{F_{3894-3910}}$$
$$CH(4300) = -2.5 \log \frac{F_{4285-4315}}{0.5F_{4240-4280} + 0.5F_{4390-4460}}$$
$$HK= 1-\frac{F_{3910-4020}}{F_{4020-4130}}.$$
A spectral index is defined in such a way as to compare the counts in a window centered on 
the molecular band or atomic line we want to measure, to the counts in a comparison 
region -- not expected to be significantly affected by absorptions from these species -- which 
sample the continuum level. 
The uncertainties related to the index measure have been obtained with the expression 
derived by \citet{Vollmann06}, assuming pure photon
noise statistics in the flux measurements.
\begin{figure}
  \centering
\resizebox{\hsize}{!}{\includegraphics{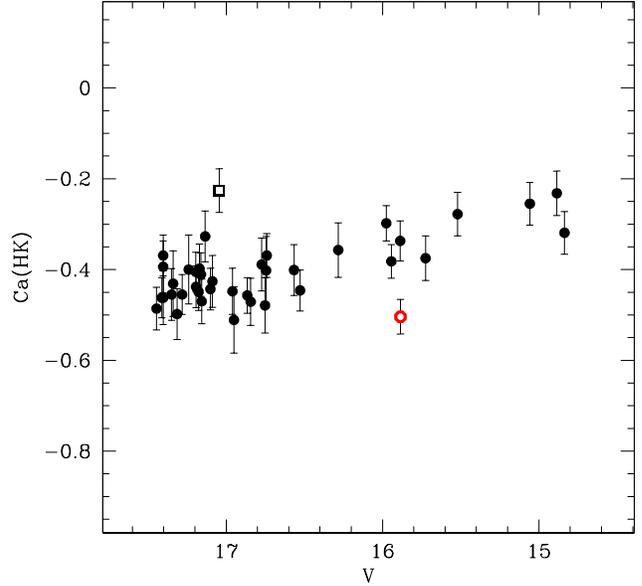}}
 
\caption{
The Ca II H and K index plotted vs. V magnitude for the M~2 giants. 
A probable non-member star is shown as an open square. 
The small scatter, fully compatible with the formal measurement errors, in the Ca(HK) values as a 
function of {\it V} provides additional evidence that all
spectroscopic targets but the notable outlier are members of M~2. 
The open red symbol refers to star 10427, which is not a member of M~2, according to its radial velocity.
}
        \label{CALCIUM}
   \end{figure}
To obtain additional membership information 
we employed the strength of the $Ca_{II}$ H and K 
lines (see Sect.~\ref{misura_indici}, as in \citealp{smith90}) as a further discriminant between cluster and field stars,
since the strength of these lines depends on the metal-abundance in this low to intermediate metallicity regime.
By assuming that M~2 is chemically homogeneous with respect to the calcium abundance, we expect that stars
belonging to the cluster show a tight sequence in the HK, $V$ plane.
We present the plot of HK index vs. the {\it V} magnitude in Fig.~\ref{CALCIUM}.
A tight relation between HK index strength and the {\it V } magnitude is clearly present for all stars
selected by using radial velocity criteria. 
From this figure, we were able to pinpoint only one outlier (ID 21729), whose spectrum has a 
noticeably strong-lined appearance. We also measured indices for this star to allow for a direct comparison
with respect to cluster members; however, we excluded this star from the abundance analysis.
Again from Fig.~\ref{CALCIUM}, we note that the probable field star (rejected according to its radial velocity), 
does occupy an anomalous position in the  the plot of HK index vs. the {\it V} magnitude.
This evidence further confirms that this stars is not a cluster member.
The measured indices, together with additional information on target stars, are listed in Table~\ref{tab_indici}.

\subsection{Index analysis}
\begin{figure}
  \centering
\resizebox{\hsize}{!}{\includegraphics{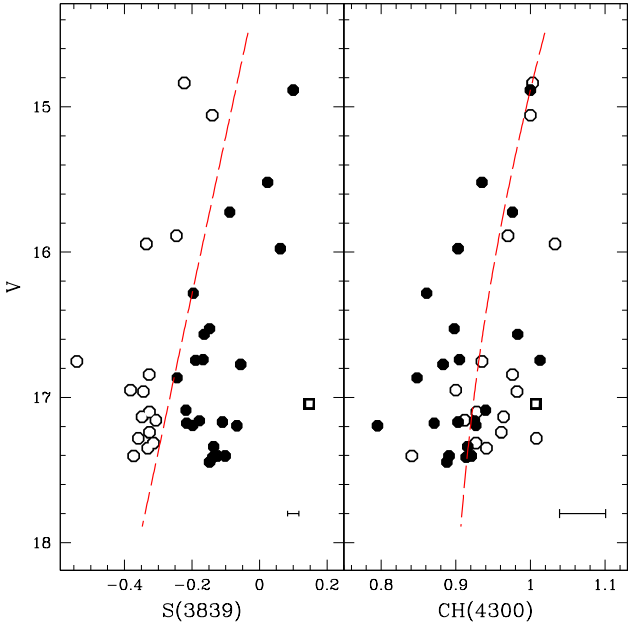}}

\caption{{\em Left panel}: Removal of gravity and temperature dependencies from CN index using median ridge line
(shown as red dashed line). Stars considered to have mid-strength or strong CN bands are depicted by 
filled circles, while CN-weak giants are represented by open circles.
A probable non-member star (see Sect.~\ref{misura_indici}) is shown as an open square. 
The median errorbar on the S(3839) and CH(4300) measurements is also shown in the 
lower-right corner of each panel.
{\em Right panel}: The same as in the left panel but for the CH index. 
The color code is consistent with the left panel.}
        \label{RIDGE}
   \end{figure}

Figure~\ref{RIDGE} shows S(3839) and CH(4300) index measurements for our data set.
Several low-resolution studies have demonstrated that the CN-band strength is a proxy for the nitrogen 
content of star atmospheres, whereas CH traces carbon \citep[e.g.,][]{smith96}.
A visual inspection of the left hand panel of Fig.~\ref{RIDGE} reveals a  clear bimodality in the CN index
over the entire magnitude range, with a few mid-strength stars.
The difference in S(3839) between CN-strong and CN-weak stars of comparable magnitude is $\sim$0.2-0.3 mag.
Giants considered to have relatively strong CN bands and CN-poor giants are represented in Fig.~\ref{RIDGE}.
The right hand panel of Fig.~\ref{RIDGE} illustrates the relation between the CN and CH band strengths for all giants:
it shows a plot of the CH(4300) index vs. the {\it V} magnitude with the CN-strong and CN-weak stars. 
In this case the spread among the measured index is very small and, in any case, within the uncertainties.
There is a tendency, as expected, for CN-strong stars to also be CH-weak, even if exceptions exist.

Out of a sample of 38 stars, 16 have weak CN bands.
The number ratio of CN-weak to CN-strong that we obtained ($\sim$0.73$\pm$0.2) is very different\footnote{
We emphasize that the ratio derived here is based on relatively few stars and the criteria for defining CN-strong stars
are different in each work.} from what is found by
\citet{smith90} (0.33; 16 RGB stars\footnote{If we exclude the two CH stars.}) and \citet{smolinski11} 
(0.35; 70 MS, SGB, and RGB stars.). 
Comparing these values directly is complicated by the fact that our study only uses RGB stars, 
while for example \citet{smolinski11} includes subgiants and dwarfs and \citet{smith90} focused on brighter stars.
Dwarfs are significantly hotter than RGB stars and less likely to show remarkable CN absorption,
thus their inclusion may bias the CN-weak to CN-strong value downward.

CN and CH bands strongly depend on the temperature and gravity, when keeping the overall abundance fixed.
These dependencies are usually minimized both
 by fitting the lower envelope of the distribution in the index-magnitude plane  
(or index-color plane; see for example \citealp{harbeck03,kayser08}) or by using median ridge lines to correct for 
the curvature introduced by both temperature and gravity effects \citep{pancino10, lardo12}.
To be consistent with our previous works, we used {\em median ridge line} (Fig.~\ref{RIDGE})
to minimize the effect of effective temperature and surface gravity in the CH and CN measurements.
The baselines adopted for the M~2 red giants to correct S(3839) and CH(4300) indices are
$$S_{0} = -0.09 \times V+1.3$$
$$CH_{0} = 0.005 \times V^{2}-0.21 \times V+ 2.88,$$
The rectified CN and CH indices are indicated as $\delta$S3839 and $\delta$CH4300, respectively, and
we refer to these new indices in the following\footnote{We obtained a rough estimate of 
the uncertainty in the placement of these median ridge lines by using the first interquartile of the rectified 
indices divided by the square root of the number of points.
The resulting uncertainties (typically $\sim$ 0.013 for the CN index and $\sim$ 0.008 for the CH index) 
are largely negligible for the applications of this work.}.
\begin{figure}
  \centering
\resizebox{\hsize}{!}{\includegraphics{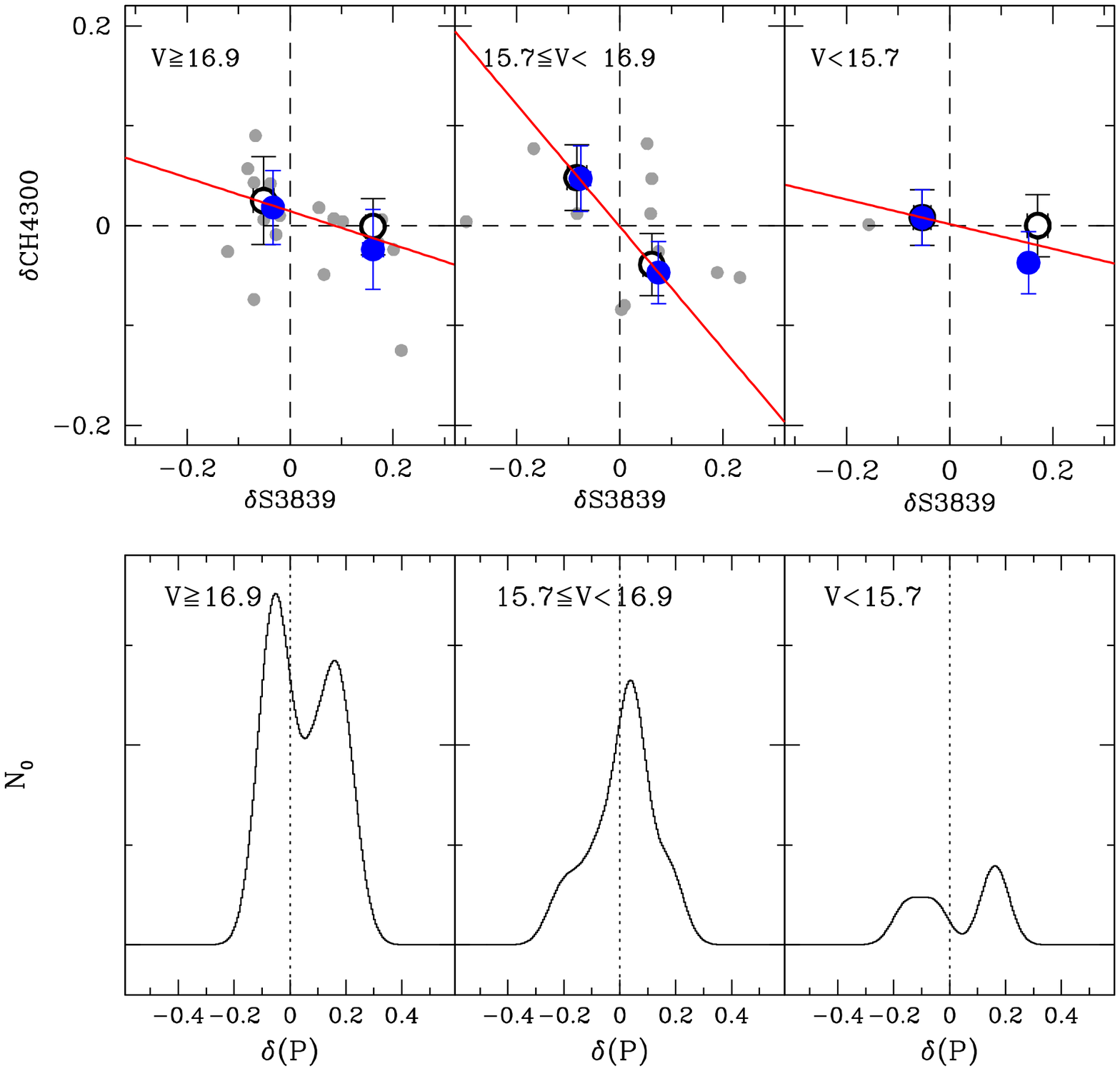}}

  \caption{{\em Upper panel}: Anticorrelation plot for the CN and CH band strengths in three magnitude bins
(V$\geq$16.9 mag, $15.7 \leq V < 16.9$ mag and $V<15.7$ mag).
Gray dots show measurements for stars. CH weak and CH strong stars are separated by the horizontal dashed line, and their centroids 
with 1$\sigma$ are marked as large white dots. CN strong and weak stars are separated by the vertical dashed line, and 
their centroids with their 1$\sigma$ are shown as large blue dots. The red continuous line connects the locus equidistant from 
CH-strong/CN-weak centroids and CH-weak/CN-strong ones.The generalized histograms in the bottom panels represent
the distribution of distances
of projected points from the origin $P$ (see text for details).}
        \label{INDICI}
   \end{figure}
\begin{table*}
\caption{Index measurements, magnitudes, colors, and radial velocities for the sample stars}             
\label{tab_indici}      
\centering          
\begin{tabular}{r c c c c c r r r r r r r r }
\hline \hline  \\

ID      &$ID_{SDSS}$ &     RA        &   Dec         &  {\it V}    &{\it U--V}&   $V_{r}$            &  CN      &  dCN     &$\delta$CN&   CH   & dCH     & $\delta$CN \\
	& 	     &    (deg)      &  (deg)        &             &          &   $km s^{-1}$        & (mag)    &  (mag)   &  (mag)   &  (mag) & (mag)   &   (mag)     \\
\hline
  1047  &   205468   &  323.35400060 &   -0.87919450 &   16.740    &  0.8360 &  $  -30	\pm 23   $    & -0.167 &  0.010   &  0.075   &  0.905 & 0.032	&  -0.026  \\
  1221  &   204057   &  323.34119890 &   -0.87748070 &   16.283    &  0.9340 &  $  -9	\pm 43   $    & -0.196 &  0.015   &  0.003   &  0.861 & 0.022	&  -0.084  \\
  1249  &   205374   &  323.35306530 &   -0.87727650 &   16.566    &  0.8020 &  $  11	\pm 37   $    & -0.164 &  0.025   &  0.062   &  0.983 & 0.039	&   0.047  \\
  1921  &   204613   &  323.34594560 &   -0.87168230 &   15.725    &  1.0500 &  $  -9	\pm 25   $    & -0.088 &  0.017   &  0.060   &  0.976 & 0.024	&   0.012  \\
  1927  &   206299   &  323.36152960 &   -0.87163050 &   17.403    &  0.7582 &  $  17	\pm 62   $    & -0.373 &  0.025   & -0.070   &  0.841 & 0.041	&  -0.074  \\
  2288  &   205760   &  323.35679530 &   -0.86887270 &   15.887    &  1.0400 &  $  -33  \pm 21   $    & -0.246 &  0.019   & -0.083   &  0.970 & 0.025	&   0.012  \\
  3190  &   206918   &  323.36713180 &   -0.86275460 &   17.177    &  0.6840 &  $  -1	\pm 22   $    & -0.216 &  0.012   &  0.066   &  0.871 & 0.022	&  -0.049  \\
  3285  &   204469   &  323.34467810 &   -0.86211390 &   16.745    &  0.7995 &  $  11	\pm 28   $    & -0.189 &  0.035   &  0.053   &  1.013 & 0.041	&   0.082  \\
  3397  &   206350   &  323.36209610 &   -0.86148000 &   17.170    &  0.7990 &  $  -32  \pm 36   $    & -0.110 &  0.009   &  0.171   &  0.903 & 0.031	&  -0.017  \\
  3760  &   205848   &  323.35765450 &   -0.85923620 &   17.161    &  0.7712 &  $  -3	\pm 25   $    & -0.178 &  0.012   &  0.102   &  0.925 & 0.028	&   0.004  \\
  4144  &   206472   &  323.36324840 &   -0.85704770 &   16.527    &  0.8623 &  $  -33  \pm 30   $    & -0.148 &  0.013   & 0 .074   &  0.898 & 0.033	&  -0.039  \\
  5010  &   206002   &  323.35884820 &   -0.85232980 &   15.058    &  1.2417 &  $  -37  \pm 30   $    & -0.140 &  0.015   & -0.053   &  1.000 & 0.028	&   0.008  \\
  5149  &   204376   &  323.34376850 &   -0.85158800 &   15.520    &  1.0842 &  $  11	\pm 38   $    &  0.024 &  0.010   &  0.153   &  0.935 & 0.031	&  -0.037  \\
  5185  &   205505   &  323.35424850 &   -0.85143600 &   17.445    &  0.7163 &  $  -8	\pm 29   $    & -0.148 &  0.023   &  0.159   &  0.888 & 0.031	&  -0.027  \\
  6609  &   112780   &  323.36574622 &   -0.84405523 &   16.752    &  0.9172 &  $  -5	\pm 27   $    & -0.541 &  0.012   & -0.298   &  0.935 & 0.023	&   0.004  \\
  9229  &   106190   &  323.33041650 &   -0.83108230 &   15.976    &  1.0558 &  $  -30  \pm 36   $    &  0.062 &  0.012   &  0.233   &  0.903 & 0.021	&  -0.052  \\
 10803  &   105859   &  323.32844101 &   -0.82363565 &   17.314    &  0.8108 &  $  8	\pm 68   $    & -0.314 &  0.009   & -0.020   &  0.927 & 0.050	&   0.010  \\
 11131  &   105112   &  323.32387272 &   -0.82204937 &   16.865    &  0.8347 &  $  -12  \pm 27   $    & -0.244 &  0.024   &  0.009   &  0.848 & 0.031	&  -0.080  \\
 11796  &   104710   &  323.32131086 &   -0.81884683 &   17.403    &  0.7250 &  $  -37  \pm 34   $    & -0.125 &  0.014   &  0.178   &  0.921 & 0.024	&   0.006  \\
 14343  &   103672   &  323.31346832 &   -0.80667018 &   16.950    &  0.7926 &  $  5	\pm 61   $    & -0.382 &  0.020   & -0.121   &  0.900 & 0.049	&  -0.026  \\
 15217  &   105961   &  323.32904057 &   -0.80229140 &   17.349    &  0.6652 &  $  10	\pm 64   $    & -0.331 &  0.023   & -0.033   &  0.941 & 0.041	&   0.025  \\ 
 16614  &   102869   &  323.30469421 &   -0.79511601 &   16.843    &  0.8018 &  $  -32  \pm 24   $    & -0.327 &  0.033   & -0.076   &  0.976 & 0.037	&   0.048  \\
 17116  &   104710   &  323.30401959 &   -0.79241062 &   15.856    &  0.7250 &  $  -14  \pm 21   $    & -0.102 &  0.010   &  0.058   &  0.891 & 0.024	&  -0.068  \\
 17978  &   109421   &  323.34565833 &   -0.78756009 &   17.282    &  0.7500 &  $  -28  \pm 24   $    & -0.359 &  0.016   & -0.067   &  1.008 & 0.037	&   0.090  \\
 18076  &   110938   &  323.35382365 &   -0.78697660 &   17.157    &  0.8453 &  $  15	\pm 60   $    & -0.307 &  0.020   & -0.027   &  0.912 & 0.044	&  -0.009  \\
 18369  &   108550   &  323.34180935 &   -0.78528604 &   15.944    &  0.9510 &  $  -7	\pm 25   $    & -0.335 &  0.015   & -0.167   &  1.033 & 0.033	&   0.077  \\
 18682  &   108795   &  323.34281919 &   -0.78336333 &   17.101    &  0.7351 &  $  15	\pm 36   $    & -0.326 &  0.015   & -0.051   &  0.928 & 0.033	&   0.006  \\
 19348  &   105097   &  323.32380781 &   -0.77915460 &   17.240    &  0.6913 &  $  -36  \pm 29   $    & -0.326 &  0.015   & -0.038   &  0.961 & 0.034	&   0.042  \\
 19928  &   110282   &  323.35007852 &   -0.77513187 &   16.773    &  0.7985 &  $  -5	\pm 25   $    & -0.056 &  0.018   &  0.189   &  0.883 & 0.028	&  -0.047  \\
 20163  &   112239   &  323.36202881 &   -0.77336422 &   17.339    &  0.8046 &  $  15	\pm 44   $    & -0.136 &  0.022   &  0.161   &  0.916 & 0.045	&  -0.001  \\
 20473  &   112049   &  323.36081815 &   -0.77080365 &   17.193    &  0.7606 &  $  -33  \pm 22   $    & -0.198 &  0.011   &  0.085   &  0.927 & 0.028	&   0.007  \\
 20654  &   113091   &  323.36801961 &   -0.76926014 &   17.412    &  0.6326 &  $  -1	\pm 25   $    & -0.138 &  0.026   &  0.165   &  0.914 & 0.040	&  -0.001  \\
 20871  &   107606   &  323.33741787 &   -0.76723645 &   16.960    &  0.7788 &  $  19	\pm 39   $    & -0.344 &  0.021   & -0.082   &  0.982 & 0.053	&   0.057  \\
 20885  &   110521   &  323.35133131 &   -0.76713157 &   17.195    &  0.7770 &  $  -28  \pm 24   $    & -0.067 &  0.015   &  0.216   &  0.795 & 0.027	&  -0.125  \\
 21053  &   113761   &  323.37197589 &   -0.76537623 &   14.885    &  1.3645 &  $  -7	\pm 48   $    &  0.100 &  0.020   &  0.171   &  1.000 & 0.031	&   0.000  \\
 21729  &   104749   &  323.32152639 &   -0.75746580 &   17.044    &  0.6698 &  $  -30  \pm 42   $    &  0.148 &  0.018   &  0.418   &  1.007 & 0.027	&   0.084  \\
 22047  &   113079   &  323.36788890 &   -0.75307288 &   14.836    &  1.4040 &  $  -33  \pm 28   $    & -0.223 &  0.017   & -0.157   &  1.003 & 0.026	&   0.001  \\
 22092  &   113393   &  323.36985251 &   -0.75241609 &   17.134    &  0.7180 &  $  -9	\pm 20   $    & -0.348 &  0.019   & -0.070   &  0.964 & 0.050	&   0.043  \\
 22170  &   113230   &  323.36891078 &   -0.75140570 &   17.088    &  0.7840 &  $  14	\pm 54   $    & -0.218 &  0.042   &  0.056   &  0.940 & 0.052	&   0.018  \\

\hline                  
\end{tabular}
\tablefoot{$ID_{SDSS}$ is the DAOPHOT ID number from \citet{an08} photometric catalog.}
\end{table*} 
Figure~\ref{INDICI} shows the rectified index  $\delta$S3839 as a function of $\delta$CH4300 for all the stars studied 
in this paper.
Abundance analysis in Sect.~\ref{risultati} confirmed that carbon abundance depends on the evolutionary state 
and decreases towards brighter luminosities.
Therefore, we separately considered stars in three different magnitude bins: $V\geq16.9$, $15.7\leq V<16.9$, and $V<15.7$ mag,
to minimize the impact of evolutionary effects on our index analysis.
To better visualize the hidden substructure in the $\delta$S3839 vs. $\delta$CH4300 plane we adopted the
method described below.
\begin{itemize}
 \item A median is used to compute the centroids of the CH-strong ($\delta$CH4300 $>$ 0) and CH-weak 
($\delta$CH4300 $<$ 0) in the CH-CN plane.
  The resulting centroids with their 1$\sigma$ errors are reported in Fig.~\ref{INDICI} along with the 
measurements for each star. We also divided also the stars into CN-strong ($\delta$S3839 $>$ 0) and CH-weak 
($\delta$S3839 $<$ 0) groups and their centroids with relative error bars are plotted in the same figure;

\item A line passing through the midpoint connecting CH-strong/CN-weak and CH-weak/CN-strong centroids is traced;

\item Each observed point in the CN-CH plane is projected onto this line;

\item We take as origin ($P$) the intersection between this line and the perpendicular line passing through the point 
($\delta$S3839, $\delta$CH4300)=(0,0);

\item A generalized histogram of the distribution of distances of projected points from the origin $P$
is constructed. 

\end{itemize}
The histograms are shown in the bottom panels of Fig.~\ref{INDICI}, where different panels show different subsamples of RGB stars.
Each data point in this histogram has been replaced by a Gaussian of unit area and standard deviation $\sigma$=0.04
\footnote{The 0.04 magnitudes used as the Gaussian width in the generalized histograms of Fig.~\ref{INDICI} is the same as
the measurement error on $\delta$S3839.}.
We distinguish between CN-strong (CH-weak) stars and CN-weak (CH-strong) stars by cutting at zero the histogram of
distances distribution. The dimension of the subsamples, and the number of CN-strong stars in each bin is listed 
in the second and third columns, respectively, in Table~\ref{PK}. 
\begin{table}
\caption{Dimension of the samples and results of KS test.}             
\label{PK}      
\centering          
\begin{tabular}{l c c l}
\hline \hline  \\

MAG BIN           & $N_{Stars}$    &    CN-s(CH-w)    &  $P_{KS}$\\
	          &	           &  	              &          \\
\hline
	          &	           &  	              &                  \\
$V \geq 16.9$ mag & 21             & 11               &$1.14e^{-0.5}$ \\
$16.9 < V \leq15 .7$ mag & 13	           &	9             & 0.002\\	
 $V < 15.7$ mag   & 4              & 2                & 0.1 \\
\hline                  
\end{tabular}

\end{table}
Figure~\ref{INDICI} shows that stars fainter than V$\simeq$16.9 display clear bimodality,
with both CN-strong (CH-weak) and CN-weak (CH-strong) stars, as is 
common among GCs of intermediate metallicity. 
For brighter giants, the distribution of the projected points
is still not described well by a single symmetric Gaussian curve: indeed, a two-sided Kolgomorov-Smirnov 
returns a probability of $P_{KS} =0.002$
($P_{KS} =1.14\times 10^{-5}$ for stars in the first magnitude bin, see the last column in Table~\ref{PK}) 
that the CN-strong (CH-weak) and CN-weak (CH-strong)
are drawn from the same parent population.
When analyzing all data sets, the underlying bimodality can be confused by evolutionary
effects (mixing), but a wide spread with three notable peaks is still present.
Again from Fig.~\ref{INDICI} (top panel), for all magnitude bins we report a clear CH-CN anticorrelation
for all magnitude bins.

\section{Abundance analysis}\label{analisi_abbondanze}
\subsection{Atmospheric parameters}
We derived stellar parameters from photometry.
The effective temperature, $T_{eff}$ , 
was calculated using \citet{alonso99}
$T_{eff}$-color calibrations for giant stars. 
We used the $(U-V)$ color from 
DOLORES photometry (once calibrated on the Stetson standard field), using
$E(B-V) =0.06$  and [Fe/H] =--1.65 from the \citet{harris96} catalog (2010 edition). 
In addition, we used --- when available --- $(B-V ), (V-J ), (V-H ),$ and $(V-K )$ colors 
from \citet{lee99} and the 2MASS \citep{2mass06} photometry. 
The final  $T_{eff}$ was the mean of the individual $T_{eff}$ values from each color weighted 
by the uncertainties for each color calibration. 
The surface gravity was determined using  $T_{eff}$ , 
a distance modulus of $(m-M) _{V}$=16.05 \citep{harris96}, 
bolometric corrections BC(V) from \citet{alonso99}, and an assumed mass of 0.8 $ M_{\odot }$ \citep{berg01} . 
The microturbulent velocity was determined using the relation,
$v_{t}=-8.6 \times 10^{-4} T_{eff} + 5.6$, 
adopted from the analysis by \citet{pila96} of metal-poor subgiant and giant stars 
with comparable stellar parameters. 
This method leads to an average microturbulent velocity estimate of
$v_{t} = 1.1 \pm  0.13$ km $s^{-1}$, therefore we chose to assign a reference microturbulent velocity of 
$v_{t} = 1.0$ km $s^{-1}$ to all our program stars.

An additional check to test the reliability of our chosen atmospheric parameters was performed using theoretical 
isochrones downloaded from the Dartmouth Stellar Evolution 
Database \citep{dart08}\footnote{\url{http://stellar.dartmouth.edu/models/isolf.html}}.
We chose an isochrone of 12 Gyr with standard $\alpha$-enhanced composition, and we projected our targets on the
isochrone (following a criterion of minimum distance from the isochrone points) in the $U, (U-V)$ diagram to obtain their parameters. 
The {\em median} difference in temperature between the two methods is approximately 20$\pm$4 K, while the difference in gravity
is negligible (on the order of 0.013$\pm$0.002).
By projecting our targets on the isochrone in the {\em intrinsically} broad  $U, (U-V)$ RGB, we could possibly erase 
differences in color (and thus in temperature) between spectroscopic targets;
therefore, we preferred to rely on the \citet{alonso99} parameter estimates.

The residual external uncertainties, which could result only in a shift of the zero point,
do not affect the amplitude of star-to-star variation in C and N,
since we want to measure the internal intrinsic spread of our sample of stars.
Table \ref{stellar_abundances} reports the $T_{eff}$, $\log g$ values, and their 
uncertainties used to derive C and N abundances.

\subsection{Abundances derivation} 
Abundances for a given element were derived by comparing 
synthetic spectra with observed spectra.
The C and N abundances were estimated by spectral synthesis of the 2$\Sigma$--2$\Pi$ band of CH (the G band) 
at $\sim$4310\AA~and the $UV$ CN band at 3883\AA~(including a number of CN features in the wavelength
range of 3876-3890\AA), respectively.
The synthetic spectra were generated using the local thermodynamic equilibrium (LTE) 
program MOOG \citep{sneden73}.
The atomic and molecular line lists were taken from the latest Kurucz compilation \citep{castelli04} and 
downloaded from the F. Castelli website\footnote{\url{http://wwwuser.oat.ts.astro.it/castelli/linelists.html}}.

Model atmospheres were calculated with the ATLAS9 code, 
starting from the grid of models available on the F. Castelli website \citep{castelli03}, 
using the values of $T_{eff}$, $\log g$, and $v_{t}$ determined as explained in the previous section.
For all the models we adopted $[A/H]=-1.5$, according to the metallicity of the cluster.
The ATLAS9 models we employed were computed with the new set of opacity distribution functions
\citep{castelli03} and excluded approximate overshooting in 
calculating the convective flux.
For the CH transitions, the $\log g_{f}$ obtained from the Kurucz database were revised downward by
0.3 dex to better reproduce the solar-flux spectrum by \citet{neckel84}
with the C abundance by \citet{caffau11}, as extensively discussed in \citet{mucciarelli12}.
\begin{figure}
  \centering
\resizebox{\hsize}{!}{\includegraphics{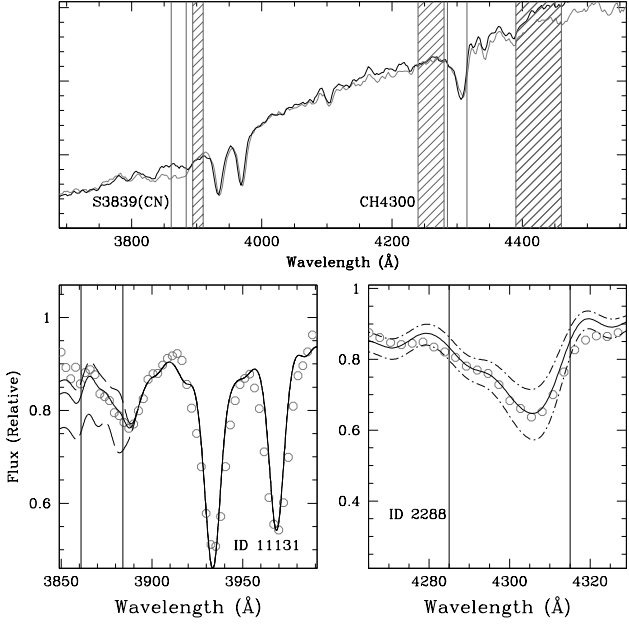}}

  \caption{{\em Top panel}:DOLORES-LRB spectra of the stars 17116 (gray) and 18369 (black) in the region of the CN $UV$ feature
and CH band. 
The stars are essentially identical in 
$V$ magnitude ($V=15.86, 15.94$, respectively), $(U-V)$ color ($(U-V)$=1.06 and 0.95, respectively),
and C abundance (see Table~\ref{stellar_abundances}). The gray shaded regions show the continuum regions, while the solid gray lines
show the window from which we measured the CN and CH indices.
{\em Bottom panels}. Observed (gray empty circles) and synthetic (line) spectra around the CN and CH band for
the stars 11131 and 2288 stars, respectively.
The solid line represents the best fit, while the dashed-dotted lines are the synthetic spectra computed with the derived C abundance 
altered by $\pm$0.2 dex and N abundance altered by $\pm$ 0.5 dex from the best value.
Vertical lines show the location of the CN (3861\AA~to 3884\AA) and CH (4285\AA~ to 4315\AA~ absorption bandpass.}
        \label{SPECTRA}
   \end{figure}
Figure~\ref{SPECTRA} illustrates the fit of synthetic spectra to the observed ones in CH and CN spectral regions.
These stars have essentially the same stellar parameters ($T_{eff} \sim$5000 $\log$ g =2.2), lying at about the 
same place in the cluster CMD, yet their CN bands differ strongly. 
Because the abundances of C and N are coupled, we iterated until self-consistent 
abundances were obtained.
Further details on the abundances derivation can be found in \citet{lardo12}. 

We assumed that all stars had the same oxygen abundance ([O/Fe]=+0.4 dex) regardless of luminosity
(constant oxygen abundance as the star evolves along the RGB).
The derived C abundance is dependent on the O abundance and therefore so is the N abundance.
In molecular equilibrium an overestimate of oxygen produces an overestimate of carbon (and vice versa), 
and an overestimate of carbon from CN features is reflected in an underestimate of nitrogen. 
We expect that the exact O values will affect the derived C abundances only negligibly,
since the CO coupling is marginal in cool stellar atmospheres (T$\leq$4500K).
To quantify the sensitivity of the C abundance on the adopted O abundance, we varied the oxygen abundances 
and repeated the spectrum synthesis to determine the exact dependence for a few representative stars 
(4900 K $\leq T_{eff} \leq$ 5400 K). 
In these computations, we adopted [O/Fe]= -0.2 dex,  [O/Fe]= 0.0 dex, and [O/Fe]= +0.4 dex.
We found that strong variations in the oxygen abundance markedly affect the derived C abundance only for the 
brighter stars in our sample, for which [C/Fe] can change by as much as 0.17-0.20 dex for a 
0.6 dex change in assumed [O/Fe].
This is within the uncertainty assigned to our measurement.
See also a discussion of the effects of 
considering different O abundance on carbon abundance derivation in \citealp{martell08}.

The total error in the derived C and N abundances was computed by taking the internal errors associated
to the chemical abundances into account. Two sources of errors can contribute to this internal error:
(i) the uncertainty introduced by errors in the atmospheric parameters used to compute chemical abundances, and
(ii) the error in the fitting procedure and  errors in the abundances that are 
likely caused by noise in the spectra.
To estimate the sensitivity of the derived abundances to the adopted atmospheric parameters,
we therefore repeated our abundance analysis and changed only one parameter at each iteration for several stars that are 
representative of the temperature and gravity range explored.

Typically, we found $\delta$A(C)/$\delta T_{eff} \simeq$ 0.09 -0.13 dex  
and $\delta$A(N)/$\delta T_{eff} \simeq$ 0.14 - 0.18 dex for the temperature.
The errors due to uncertainties on gravity and microturbulent velocity are negligible (on the order of 0.03 dex or less).
The contribution of continuum placement errors was estimated by determining the change in the abundances as the 
synthetic/observed continuum normalization was varied\footnote{We continuum-normalized our spectra 
using the same function (cubic spline) in the task IRAF {\em continuum} but with an order slightly higher with respect 
to that chosen for the first normalization.}: generally, this uncertainty added 0.11 dex to the abundances.
The errors derived from the fitting procedure  were then added in quadrature to the errors introduced by 
atmospheric parameters, resulting in 
an overall error of approximately 0.20 dex for the C abundances and 0.22 dex for the N values.

We present the abundances derived as described above and the relative uncertainties in the abundance 
determination in Table~\ref{stellar_abundances}. Additionally, this table lists the derived atmospheric 
parameters of all our targets.
\begin{table}
\caption{Atmospheric parameters and carbon and nitrogen abundances for sample stars.}             
\label{stellar_abundances}      
\centering          
\begin{tabular}{c c c c c c }
\hline \hline  \\

 ID  &      $T_{eff}$ & $dT_{eff}$   &   $\log g$  &  A(C)   &  A(N) \\
	          &	(K) & (K)                  &   (cgs)	             &   (dex) &   (dex)\\
\hline
  1047  &    5184 &  60   & $ 2.7 \pm 0.03 $ & $  5.80  \pm    0.19 $	& $ 7.50 \pm	 0.23  $ \\
  1221  &    5041 &  56   & $ 2.4 \pm 0.03 $ & $  6.11  \pm    0.19 $	& $ 6.75 \pm	 0.22  $ \\
  1249  &    5111 &  59   & $ 2.6 \pm 0.03 $ & $  6.14  \pm    0.19 $	& $ 6.30 \pm	 0.23  $ \\
  1921  &    4955 &  54   & $ 2.2 \pm 0.03 $ & $  5.92  \pm    0.16 $	& $ 7.23 \pm	 0.22  $ \\
  1927  &    5304 &  85   & $ 3.0 \pm 0.03 $ & $  6.22  \pm    0.22 $	& $ 7.25 \pm	 0.25  $ \\
  2288  &    4959 &  54   & $ 2.2 \pm 0.03 $ & $  6.07  \pm    0.18 $	& $ 7.18 \pm	 0.22  $ \\
  3190  &    5581 &  71   & $ 3.0 \pm 0.03 $ & $  6.47  \pm    0.23 $	& $ 7.48 \pm	 0.27  $ \\
  3397  &    5301 &  84   & $ 2.9 \pm 0.03 $ & $  6.28  \pm    0.20 $	& $ 7.15 \pm	 0.24  $ \\
  3760  &    5421 &  86   & $ 2.9 \pm 0.03 $ & $  6.46  \pm    0.22 $	& $ 7.44 \pm	 0.25  $ \\
  4144  &    5142 & 107   & $ 2.6 \pm 0.05 $ & $  5.87  \pm    0.21 $	& $ 7.40 \pm	 0.23  $ \\
  5010  &    4732 &  65   & $ 1.8 \pm 0.03 $ & $  5.68  \pm    0.17 $	& $ 6.96 \pm	 0.22  $ \\
  5149  &    4904 &  53   & $ 2.0 \pm 0.03 $ & $  5.59  \pm    0.20 $	& $ 7.56 \pm	 0.21  $ \\
  5185  &    5413 &  67   & $ 3.0 \pm 0.03 $ & $  6.27  \pm    0.22 $	& $ 7.06 \pm	 0.25  $ \\
  9229  &    5022 &  75   & $ 2.3 \pm 0.03 $ & $  5.87  \pm    0.21 $	& $ 7.53 \pm	 0.22  $ \\
 10803  &    5458 &  88   & $ 3.0 \pm 0.03 $ & $  6.35  \pm    0.21 $	& $ 7.12 \pm	 0.26  $ \\
 11131  &    5326 &  85   & $ 2.8 \pm 0.03 $ & $  6.18  \pm    0.20 $	& $ 7.17 \pm	 0.25  $ \\
 11796  &    5282 &  87   & $ 3.0 \pm 0.04 $ & $  5.91  \pm    0.20 $	& $ 7.77 \pm	 0.24  $ \\
 14343  &    5229 &  83   & $ 2.8 \pm 0.03 $ & $  6.12  \pm    0.20 $	& $ 7.25 \pm	 0.24  $ \\
 15217  &    5328 &  88   & $3.0 \pm  0.04 $ & $  6.24  \pm    0.20 $   & $ 6.85 \pm     0.25  $ \\
 16614  &    5224 &  83   & $ 2.7 \pm 0.02 $ & $  6.17  \pm    0.19 $	& $ 6.75 \pm	 0.24  $ \\
 17116  &    5006 &  73   & $ 2.2 \pm 0.03 $ & $  6.13  \pm    0.17 $	& $ 7.58 \pm	 0.22  $ \\
 17978  &    5263 &  85   & $ 2.9 \pm 0.03 $ & $  5.98  \pm    0.19 $	& $ 7.19 \pm	 0.24  $ \\
 18076  &    5232 &  82   & $ 2.8 \pm 0.03 $ & $  5.99  \pm    0.22 $	& $ 7.17 \pm	 0.24  $ \\
 18369  &    4956 &  73   & $ 2.2 \pm 0.03 $ & $  6.12  \pm    0.22 $	& $ 6.11 \pm	 0.22  $ \\
 18682  &    5397 &  87   & $ 2.9 \pm 0.03 $ & $  6.40  \pm    0.24 $	& $ 6.52 \pm	 0.26  $ \\
 19348  &    5383 &  87   & $ 2.9 \pm 0.03 $ & $  6.41  \pm    0.22 $	& $ 6.85 \pm	 0.25  $ \\
 19928  &    5146 &  81   & $ 2.7 \pm 0.03 $ & $  5.82  \pm    0.22 $	& $ 7.41 \pm	 0.23  $ \\
 20163  &    5179 &  82   & $ 2.9 \pm 0.04 $ & $  5.99  \pm    0.19 $	& $ 7.10 \pm	 0.24  $ \\
 20473  &    5200 &  82   & $ 2.8 \pm 0.03 $ & $  5.92  \pm    0.21 $	& $ 7.46 \pm	 0.23  $ \\
 20654  &    5488 & 125   & $ 3.0 \pm 0.04 $ & $  6.46  \pm    0.22 $	& $ 6.84 \pm	 0.26  $ \\
 20871  &    5263 &  83   & $ 2.8 \pm 0.03 $ & $  6.27  \pm    0.21 $	& $ 6.79 \pm	 0.24  $ \\
 20885  &    5076 &  80   & $ 2.8 \pm 0.04 $ & $  5.45  \pm    0.19 $	& $ 7.47 \pm	 0.23  $ \\
 21053  &    4661 &  62   & $ 1.7 \pm 0.03 $ & $  5.61  \pm    0.16 $	& $ 7.21 \pm	 0.23  $ \\
 22047  &    4630 &  61   & $ 1.6 \pm 0.03 $ & $  5.57  \pm    0.28 $	& $ 6.48 \pm	 0.24  $ \\
 22170  &    5271 &  83   & $ 2.8 \pm 0.03 $ & $  6.21  \pm    0.20 $	& $ 7.16 \pm	 0.24  $ \\
\hline                  
\end{tabular}

\end{table} 

\section {C and N abundance results}\label{risultati}
\begin{figure}
  \centering
\resizebox{\hsize}{!}{\includegraphics{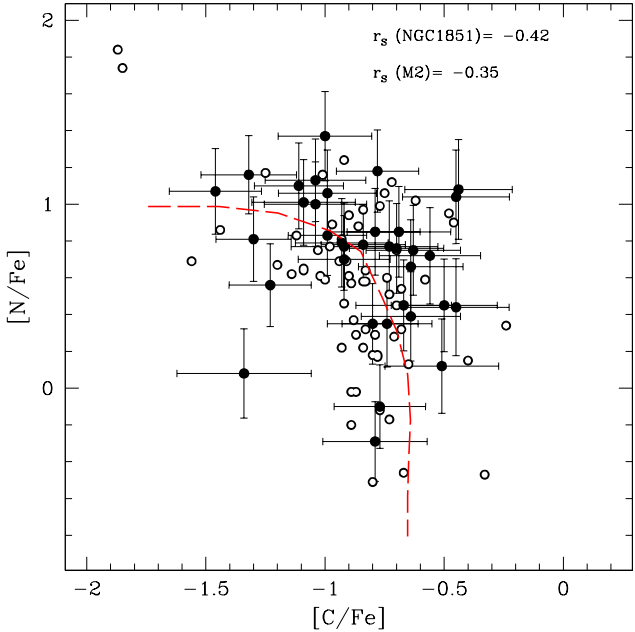}}

  \caption{Derived [N/Fe] abundances for M~2 stars in Table~\ref{stellar_abundances} as a 
function of the [C/Fe] abundances from our sample (filled circles).
A C vs. N anticorrelation is apparent. For comparison we also plotted our previous results on a sample of MS and SGB stars in the 
cluster NGC~1851 (white dots). The red dashed line indicates the relationship, shown over its full range, that prevails 
in NGC~1851 from our earlier work.}
        \label{abbondanze}
   \end{figure}

Variations in light-element abundances were already observed in all GCs studied to date, 
and are also present in M~2.
Carbon and nitrogen exhibit the typical anticorrelation, as shown in Fig.\ref{abbondanze},
where the [C/Fe] values are plotted as a function of [N/Fe] 
with their uncertainties. For three stars out of 38, we were not able to derive C and N abundances because of the low
$S/N$ in the CN band spectral region.
We observe modest variations in carbon abundances (from [C/Fe]$\sim$ --1.4 to [C/Fe]$\sim$ --0.4) 
mildly anticorrelated (Spearman's rank correlation coefficient $r_{S}^{M~2}=-0.35$) with strong variations 
in N, which span almost 2 dex, from [N/Fe]$\sim$--0.3 up to [N/Fe]$\sim$1.4 dex.
In the same figure we also plot C and N abundances derived for 
NGC~1851 in our previous work \citep{lardo12} with the [C/Fe]-[N/Fe] relationship that prevails for these stars 
($r_{S}^{NGC~1851}=-0.42$).
The range of the spread in both C and N is about the same for M~2 and NGC~1851\footnote{For comparison,
the median value of carbon abundance is [C/Fe]=--0.79 dex ($\sigma$=0.2) for M~2 and [C/Fe]=--0.84 dex ($\sigma$=0.12dex);
respectively. Median nitrogen abundances are [N/Fe]=0.77 ($\sigma=$0.31 dex) for M~2 and [N/Fe]=0.61 for NGC~1851 ($\sigma=$0.30).}
\citep[and fully agrees with the C and N abundances presented by][for M~71, 47~Tuc, M~5, M~13, and M~15]{cohen05}.
The two anticorrelations clearly follow a similar overall pattern in the [C/Fe] vs [N/Fe] plane.

\subsection{Evolutionary effects}\label{effetti_evolutivi}
\begin{figure}
\centering
\resizebox{\hsize}{!}{\includegraphics{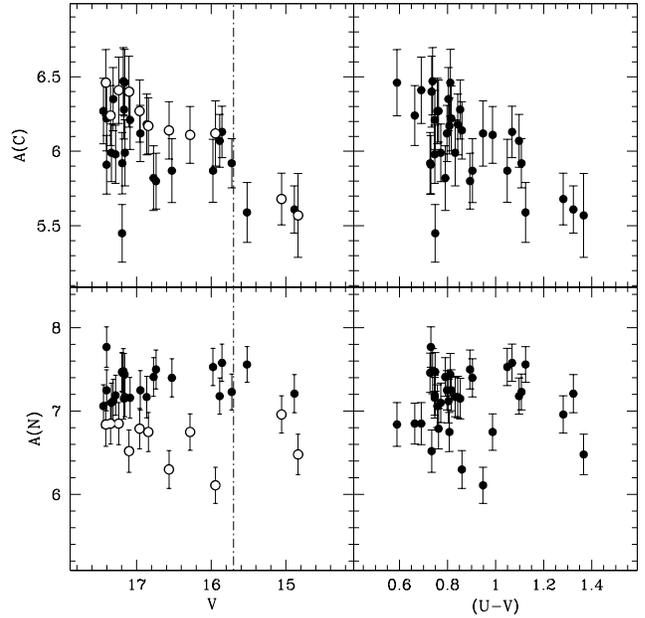}}
\caption{Derived C and N abundances plotted against the {\it V} magnitude and $U-V$ color for M~2 giants. The dot-dashed
lines indicate the luminosity at which the RGB bump occurs ($V\sim$15.7 mag).
Relatively N-rich and N-poor stars are shown in the left panels as filled and open symbols, respectively.}
        \label{TCOL}
   \end{figure}

\begin{figure}
\centering
\resizebox{\hsize}{!}{\includegraphics{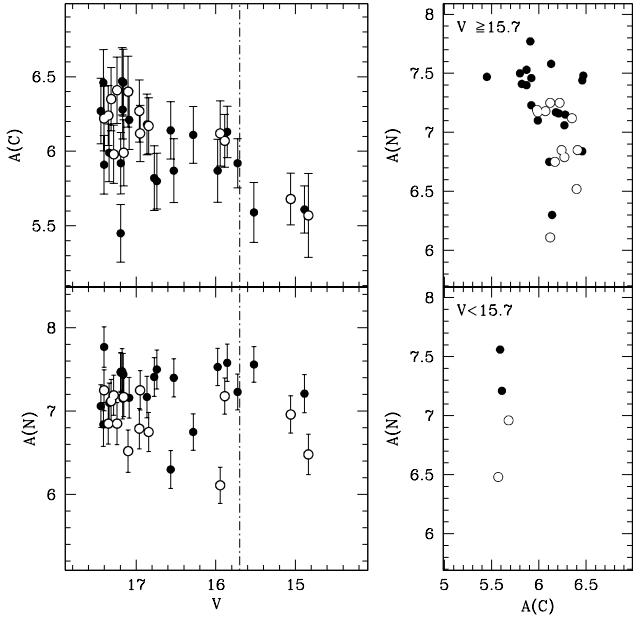}}
\caption{ {\em Left panels}: derived C and N abundances are plotted against the {\it V} magnitude for M~2 RGB stars. The dot-dashed
vertical lines indicate the luminosity at which the RGB bump occurs ($V\sim$15.7 mag).
CN-strong and CN-weak stars, as defined in Sect.~\ref{misura_indici}, are shown as filled and open symbols, respectively.
{\em Right panel}: CN-strong and CN-weak stars (see Sect.~\ref{misura_indici}) are plotted in the A(C)vs A(N) plane for
stars with luminosities fainter ({\em top}) and brighter ({\em bottom}) than the RGB bump.
The color code is consistent with the left panel.}
        \label{CONFRONTO}
   \end{figure}

As described in the Introduction, surface abundance changes due to deep mixing are not expected to occur 
in stars fainter than the RGB bump.
We plotted the derived abundances as a function of the $V$ magnitude and $U-V$
color in Fig.~\ref{TCOL} to evaluate possible systematic effects with luminosity and temperature.
While none of these effects is apparent, the top panel of Fig.~\ref{TCOL} again illustrates the notable depletion in the carbon
abundances with luminosity \citep[][and references therein]{smith03,gratton00}.
The surface carbon abundance depletion along the RGB of M~2 can be straightforwardly interpreted within a deep-mixing framework.
This implies that some form of deep mixing (i. e., meridional circulation currents, turbolent diffusion or some
similar processes), which extends below the base of the conventional convective zone,
must circulate material from the base of the convective envelope down into the CN(O)-burning region 
near the hydrogen-burning shell.
The onset of the decline in the carbon abundance appears from Fig.~\ref{TCOL} to occur at magnitude $V\simeq15.7$: 
the strong C decline for stars brighter than $V\lesssim$15.7 can be interpreted as 
the signature of the extra mixing common among metal-poor cluster giants as they cross the RGB bump.
Restricting our sample to those giants fainter than the RGB bump, we found an average C abundance of  A(C)=$6.11 \pm 0.23$.
A significant decrease in C abundance occurs at
about $V\lesssim$ 15.7, which is essentially the location of the RGB bump in this cluster 
\citep[$V_{BUMP}\sim$15.82$\pm$0.05,][]{dicecco10}: the average value for this group of upper RGB stars
is  A(C)=$5.61 \pm 0.05$ dex. 
Naturally, the extent of the carbon (nitrogen) depletion (enhancement) depends on the value of [O/Fe]
used in the analysis. 
For comparison, in metal-poor field giants \citep{gratton00},
a drop in the surface \element[][12]{C} abundance by about a factor 2.5, is seen after this second mixing episode. 
To connect CN index measurements with carbon and nitrogen abundances derived 
by spectral synthesis, we labeled CN-strong and CN-weak stars in Fig.~\ref{CONFRONTO} as defined in Sect.~\ref{misura_indici} in the 
A(C) and A(N) vs. V mag and A(C) vs. A(N) planes. 
From Fig.~\ref{CONFRONTO}, we note good agreement between the underlying [N/Fe] abundance and the measured CN band strength:
as expected CN-strong and CN-weak stars tend to occupy two separate regions in the A(C)-A(N) diagram.
Any difference of [C/Fe] at a given magnitude is difficult to interpret since it can arise from systematic differences between 
the analysis techniques.
As discussed in \citet{smith03}, a reasonable estimate of the dependence of the carbon abundance on luminosity
can be obtained by deriving $d[C/Fe]/dM_{V}$.
To compare the behavior of [C/Fe] among from field giants with M~2 giants, we fit a linear
least-squares regression of [C/Fe] against $M_{V}$ for stars with --0.8 $\leq M_{V} \leq$ 1.6.
We restricted our attention to stars selected by \citet{smith03}\footnote{We consider the the restricted sample
with the exclusion of stars HD97 and HD218857.} from the \citet{gratton00} survey.
In close analogy with \citet{smith03}, we limited our fit to stars with $M_{V} \leq 1.6$, because there is only
a slight variation below this luminosity level \citep[see Fig.~10 of][]{gratton00}.
The upper limit in luminosity was chosen to compare only the overlapping region between the two data sets.
As far as can be ascertained from the carbon abundances, the rate of mixing in this cluster is comparable to 
the one for halo field stars and many cluster giants. We found a dependence of $d[C/Fe]/dM_{V}$=0.21$\pm$0.16 that is very 
similar within the observational errors to that found
among halo field giants ($d[C/Fe]/dM_{V}$=0.20$\pm$0.03 dex) and other GCs
\citep[e.g., M~3, NGC~6397, and M~13;][]{smith03}.

From the bottom left hand panel of Fig.~\ref{TCOL}, we see no significant trend in the N
abundance with either luminosity or color: the average nitrogen abundance we found for stars fainter than the RGB bump 
(A(N)=$7.1\pm 0.4$) agrees within the quite large error bar with the one obtained
for the more luminous stars after the LF bump (A(N)=$6.9\pm 0.6$).
We tentatively divided the target stars between candidate first-generation and candidate second-generation 
(N-poor and N-rich component, respectively) stars by adopting a threshold in nitrogen abundance A(N)=7.0. In
Fig.~\ref{TCOL}, N-poor and N-rich stars are plotted, where
 we note that N-rich stars are systematically C-poor and vice versa, to further
support the presence of C-N anticorrelation.
Finally, \citet{gratton00} show an abrupt increase in N abundance of about $\sim$ 4 at $\sim V_{BUMP}$ for field giants.
Here we could not detect such a trend as the effect of the poor statistics (4 stars)
towards higher luminosities.

\subsection{C-N anticorrelation}\label{sezione_anticorrelazione}
We have seen in Sect.~\ref{effetti_evolutivi} how deep mixing affects nitrogen and (strongly) carbon abundances, because it 
introduces carbon-depleted material into the stellar convective envelopes.
All our target stars have luminosities well above the first dredge-up onset, so we expect that
their atmospheres are already depleted in carbon abundance\footnote{Among
the field stars, \citet{gratton00} data support the occurrence of a small ($\approx$ 0.1 dex) drop in the region 
of the first dredge-up.}.
A matter we plan to investigate now is how to disentangle the {\em intrinsic} star-to-star differences in surface carbon and
nitrogen abundances from the changes resulting from normal stellar evolution.
\begin{figure}
  \centering
\resizebox{\hsize}{!}{\includegraphics{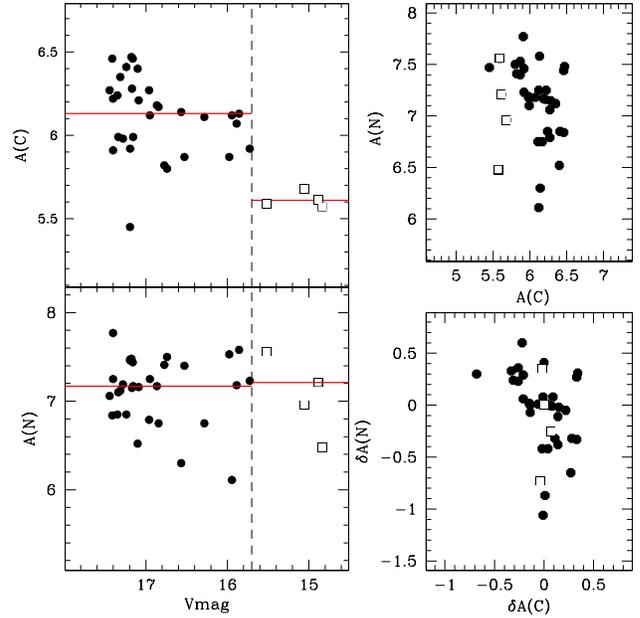}}

  \caption{Correction of the C-N anticorrelation for evolutionary effect. The left panels show the runs of A(C) and A(N)
vs. $V$ magnitude for spectroscopic targets. The vertical dashed line marks the RGB bump position.
The red continuous red line indicates {\em median} value of the carbon and nitrogen abundance for stars in two bins of magnitude
($V<15.7$ and $V\geq15.7$ mag). The top right panel shows the derived C-N anticorrelation uncorrected for carbon decline 
(and nitrogen enhancement) due to evolution of the stars along the RGB. The bottom right panel shows the {\em corrected} C-N anticorrelation. In this case we plotted
the difference of A(C) and A(N) from the {\em median} abundance value shown in the left panels (see text for further details).}
        \label{TREND}
   \end{figure}
First, we note that we cannot arbitrarily distinguish between two groups of stars with different
A(C) or A(N) for stars fainter than the LF peak, because we are unable to detect any clear bimodality.
To make more quantitative statements about bimodality, a KMM test \citep{ashman94}
was applied to the data\footnote{The star 22047 with an anomalously low 
carbon abundance ( A(C) $\simeq$ 5.6) is excluded from the fit.}.
Under the assumption that the two Gaussians have the same dispersion (homoscedastic test),
we can confirm that there is no bimodality in either A(C) or A(N) for stars with $V\geq$ 15.7.
At this point we proceed to analyze the C-N anticorrelation as follows:
\begin{itemize}
 \item computed the median abundance of carbon and nitrogen for stars with $V$ magnitude $< 15.7$ and $\geq 15.7$ mag
 (traced in red in Fig.~\ref{TREND});
\item for each measured point in the A(C)-A(N) vs. $V$ magnitude plane,  and calculated the difference between A(C), A(N) 
  and the {\em median} carbon and nitrogen abundance ($\delta A(C)$ and $\delta A(N)$, respectively);
\item constructed a plot of the $\delta A(C)$ vs. $\delta A(N)$.
\end{itemize}
The {\em corrected} $\delta A(C)$ vs. $\delta A(N)$ anticorrelation is shown in the bottom right hand 
panel of Fig.~\ref{TREND}. In this case, having corrected for the carbon decline due to normal stellar evolution,
the anticorrelation appears tighter ($r_{S}^{M~2~corrected}=-0.43$).
\begin{figure}
  \centering
\resizebox{\hsize}{!}{\includegraphics{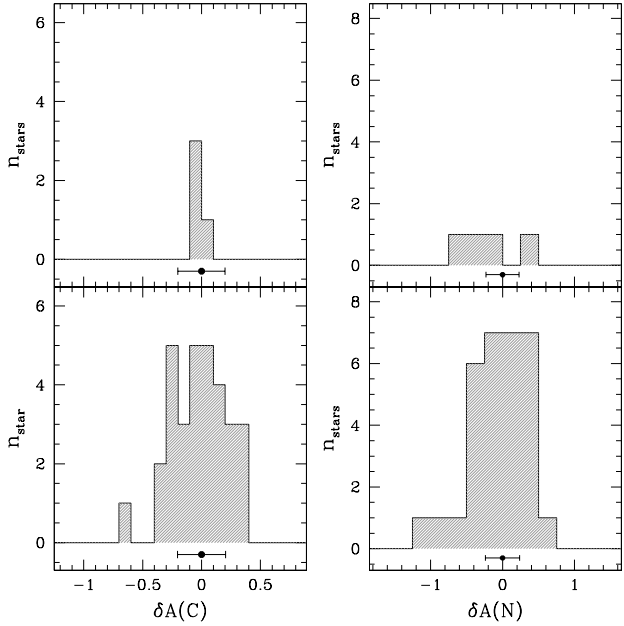}}

  \caption{Histograms of the $\delta$A(C) and $\delta(N)$ (see text) distributions. 
The two vertical panels correspond to two bins of $V$ mag ($V \geq 15.7$ and $V< 15.7$ mag;
from bottom to up).The bin size is set to 0.1 and 0.25 for
the $\delta$A(C) and $\delta(N)$, respectively. Typical median error bars  are plotted below each histogram.}
        \label{ISTOGRAMMI}
   \end{figure}

\begin{figure}
  \centering
\resizebox{\hsize}{!}{\includegraphics{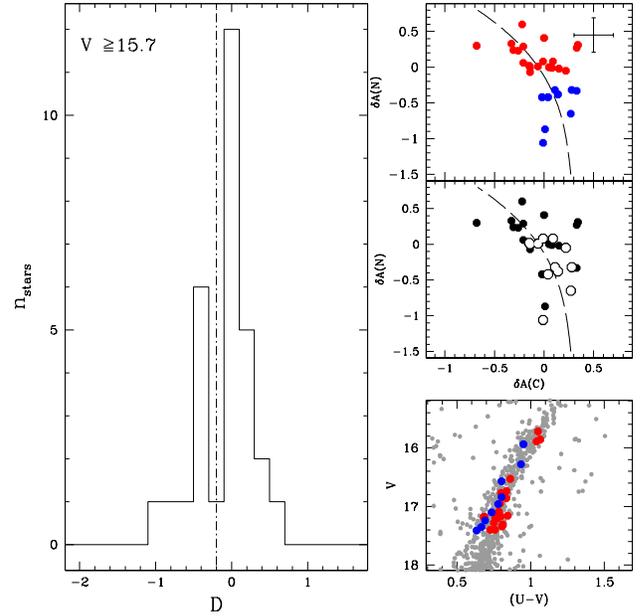}}

  \caption{{\em Left panel}: distribution of the projected distance D of stars with $V\geq 16.9$ on the fiducial plotted
in the top right panel. The dotted-dashed line separates stars belonging to the two different N groups.
{\em Top right}: {\em corrected} C-N anticorrelation for stars
with  $V\geq 15.7$. We {\em tentatively} discriminated between first (blue) and second (red) populations. The error bar
represents the typical errors on A(C) and A(N). {\em Middle right:} CN-strong and CN-weak stars are plotted in the 
$\delta$A(C)-$\delta$A(N) plane. {\em Bottom right:} first- and second-generation stars are superimposed to $V, (U-V)$ CMD of M~2.
The color code is used consistently in each panel.}
        \label{FAINT}
   \end{figure}

To better visualize the distribution of {\em corrected} C and N abundance in the two magnitude bins,
we constructed histograms of the $\delta A(C)$ and $\delta A(N)$ distribution in Fig.~\ref{ISTOGRAMMI}. 
For stars below the RGB bump we note hints of bimodality in $\delta A(C)$.
Despite the low statistics, the {\em corrected} C-N anticorrelation shows evidence for
bimodality in the distribution of N abundances, with at least two (or three) 
groups of stars populating the extremes of high N or low N (see the lower right panel of Fig.~\ref{TREND}).
To consider stars in the same evolutionary stage as much as possible, 
we first focused on the the {\em corrected} C-N anticorrelation for the faintest stars in our sample
 with magnitudes below the RGB bump ($V \geq$ 15.7).
To confirm this suggestion, we analyzed the {\em corrected} distribution of stars along the C-N anticorrelation 
using a procedure similar to the one described in \citet{marino08}.
In brief, we first draw a fiducial (shown in the top right panel of Fig.~\ref{FAINT}) 
by putting a best-fit spline through the median abundance found in successive 
short intervals of $\delta$ A(N). Then we projected each program star in the $\delta$A(C)-$\delta$A(N) anticorrelation 
on this fiducial and plotted the histogram of the distribution of vertical distances (D) of the  projected points
from the line  $\delta$A(N)=0.
The histogram is shown in the left panel of Fig.~\ref{FAINT}.
In this case (at least) two substructures are apparent, peaked at $\simeq$ --0.4 and 0.2.
We tentatively divided RGB stars between a candidate first generation and a candidate second generation by
setting an arbitrary separation at $D=-0.2$.
To allow a direct comparison between CN-strong (as derived in Sect.~\ref{INDICI}) and these second-generation 
stars, we plotted CN-strong stars in the $\delta$A(C)-$\delta$A(N) plane in the same figure.
We note that the smearing of CN-strong and CN-weak stars that happens in the A(C)-A(N) plane (see Fig.~\ref{TCOL}) is still 
present in the $\delta$A(C) vs. $\delta$A(N) plot.

A visual inspection of Fig.~\ref{FAINT} suggests that the extent of the C-N anticorrelation in the 
{\em projected} plane for second-generation (Na-N/rich) stars is greater than the errors associated with 
abundance measurements. This evidence possibly suggests the presence of a third group of stars\footnote{Stars with 
E (Extreme)  composition, by adopting the nomenclature first introduced by \citet{carretta2009}.};
unfortunately, because of uncertainties on abundance measurements and low statistics,  we cannot 
provide conclusive evidence.

In general, when stars with available Na and O abundances have been identified in the {\it U} vs. ({\it U--B}) CMD 
(or in a different color combination that includes the blue filters), it was found that the group
of Na-poor stars are systematically spread on the blue side of the RGB, while the Na-rich
population define a narrow sequence on the red RGB \citep{gru01,marino08,han09,milone10}.
Several authors have demonstrated that a clear correlation exists between N abundances (and
so $\lambda$3883 CN band strength) and Na, O and Al abundances \citep[see for example][and references therein]{marino08}.
N-rich (CN-strong) stars clearly show significantly enhanced Na abundance.
In contrast, N-poor (CN-weak) stars have a higher O content than the N-rich ones.

In the bottom right hand panel of Fig.~\ref{FAINT}, N-rich and N-poor stars are superimposed on M~2 {\it V, (U-V)} DOLORES CMD.
N-poor and N-rich stars are clearly separated into two parallel sequences in the broader giant branch seen in 
the $V, (U-V)$ diagram, with the N-rich stars systematically  appearing redder than N-poor ones, 
a behavior strictly analogous to what is observed \citet{marino08} in M~4.
It is clear that the strength of the CN and NH bands strongly influences the $U-V$ color, 
the NH band around 3360\AA, and the CN bands around 3590, 3883, and 4215\AA, located in the $U$ 
being the main contributors to the effect \citep[see also][]{sbordone2011}.

\section{The anomalous RGB  in M~2}\label{rgb_anomalo}
As discussed in Sect.~\ref{introduzione}, GCs are essentially monometallic, e.g., all the stars in 
a cluster show the same [Fe/H] abundance.
Besides the remarkable exception of $\omega$ Centauri \citep[see][and references therein]{marinOmega},
variations in the heavy element content have been detected only for a few clusters:
M~22 \citep{marino12}, Terzan~5 \citep{ferraro09, origlia11}, M~54 \citep{carretta54},
and NGC~1851 \citep{yong08, carretta10}.
In particular, among the clusters that displayed this anomalous behavior, NGC~1851 and M~22 appear rather peculiar.
For these clusters, a bimodal distribution of $s$-process elements abundance has been identified
\citep{yong08, marino12}. The chemical inhomogeneity reflects itself in a complex CMD: multiple stellar groups 
in M~22 and NGC~1851 are also clearly manifested by a split in the SGB region \citep{piotto09,milone08} which appears to be related 
to chemical variations observed among RGB stars \citep{marino12,lardo12}.
Indeed, carefully constructed CMDs ---based on colors that include a blue filter \citep[][]{han09,lardo12,marino12}---
clearly reveal that the bright SGB is connected to the blue RGB, while red RGB stars are linked to the faint SGB.
The split of the RGB discovered in the $U-I$ and $U-V$ colors for NGC~1851 and M~22, respectively,
would not be detected in the usual optical colors.

M~2 DOLORES photometry (see Fig.~\ref{M2-DOLORES}) displays an {\em anomalous} branch beyond the red edge of the
main body of the RGB. 
The difference in color between stars belonging to this structure and {\em normal} RGB stars 
is quite large (on the order of 0.2-0.3 mags, well above the typical measurement errors) and extends down to the SGB region.
There may be a second group of stars that are 0.3 mags redder with respect to this sequence and can possibly 
be more, the anomalous RGB stars.
Unfortunately, because of low statistics, we cannot provide a conclusive evidence and radial velocity and proper motion measurements
should be made to see whether these stars are members of the cluster.
 \begin{figure}
  \centering
  \resizebox{\hsize}{!}{\includegraphics{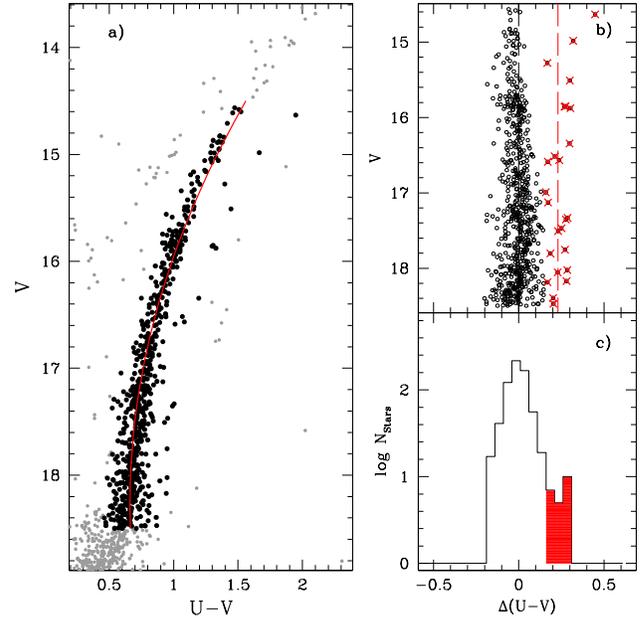}}
\caption{(a) $U, V$ CMD from DOLORES images is shown in gray. Selected RGB stars are plotted as black dots, while the red continuous line
is the fiducial obtained in the way described in the text. Panels (b) and (c) show the rectified RGB in function of the color difference
and the histogram color distribution, respectively (see the text for details).}
        \label{DOLO}
   \end{figure}

As a high-latitude system, M~2 is not affected by high interstellar absorption ($E(B-V)$=0.06; 
\citealp{harris96}, 2010 edition), and it is very unlikely that the differential reddening has caused the double RGBs. 
The color difference between the two RGBs in the $U-V$ color, at the given $V$ magnitude of the horizontal-branch (HB) level,
is $\sim$0.3 mag, which is about three times more than the maximum color difference expected in the extreme situation 
where one group of stars is all reddened by $E(B-V)$ = 0.06, while the other group has $E(B-V)$ = 0.00. 
Because the additional RGB sequence only amounts to a small fraction of the 
total giant population, we cannot exclude field contamination as the cause of the observed additional RGB branch.
We expect a very modest degree of contamination 
by Galactic fore/background stars, because of the combination between the relatively high (absolute) 
Galactic latitude of the cluster (b=-- 36\degr) and the small area of the considered annular field (1\arcmin $<R<$4\arcmin).
We used the Galactic model TRILEGAL \citep{girardi05}\footnote{\url{ http://stev.oapd.inaf.it/cgi-bin/trilegal}}
to obtain a conservative estimate of the degree of contamination affecting the samples of candidate RGB stars 
with 0.4$\leq(U-V)\leq$2 and 18.5$\leq V \leq$14.5 mag in the present analysis (Fig.~\ref{DOLO}).
We found that the fraction of Galactic field stars in our samples is lower than 1\% in the considered annular field.

\begin{figure}
  \centering
  \resizebox{\hsize}{!}{\includegraphics{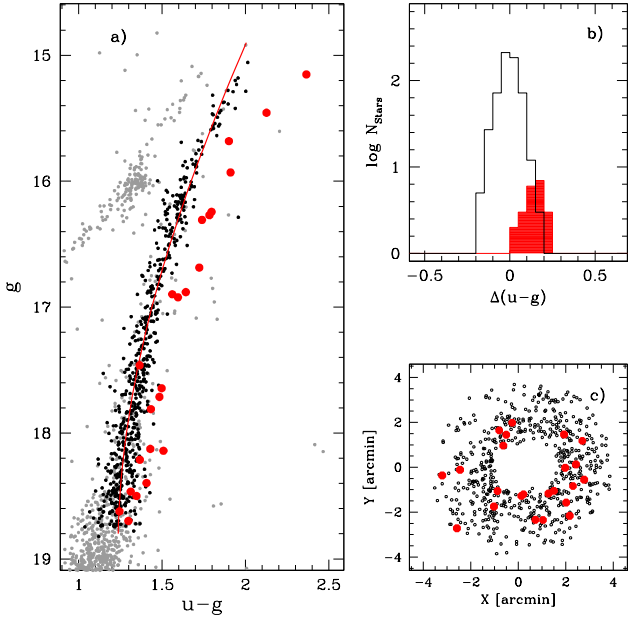}}
\caption{(a) $u, (u-g)$ CMD from \citet{an08} {\em corrected} photometry zoomed in around the RGB. Stars selected as red in Fig.~\ref{DOLO} are 
plotted as red circles, while the red continuous line is the fiducial obtained in the way described in the text. Panels (b) 
show the color distribution in the $(u-g)$ color, while panel (c) shows spatial distribution of the selected red stars.}
        \label{CHECK}
   \end{figure}

To take photometric errors into due account, we follow the method described in \citet{anderson09} 
to distinguish intrinsic color broadening from unphysical photometric error effects.
We considered the two independent CMDs obtained from DOLORES and \citet{an08} photometry.
In Fig.~\ref{DOLO} we selected the portion of the RGB sequence with magnitudes between $14.5 \leq V \leq 18.5$ mag.
In addition, we defined {\em bona fide} RGB members as the stars closer to the main RGB locus in the corrected DOLORES CMD
(panel (a) of Fig~\ref{DOLO}). We obtained the RGB fiducial as described in \citet{milone08}.
In brief, we drew a ridge line (fiducial) by putting a best-fit spline through the average color computed
in successive short (0.2 mag) magnitude intervals. In panel (b) we have subtracted from the color of each star the color of the fiducial at the same
magnitude and plotted the $V$ magnitude in function of this color difference; $\Delta (U-V)$.
The histogram color distribution on a logarithmic scale in panel (c) presents a clear substructure at the red end of the RGB, and we 
arbitrarily isolated RGB stars with $\Delta (U-V)>$0.15. These stars are plotted in panel (b).
If the red branch we see is due to photometric errors, then a star redder than the RGB ridge line in the $V, (U-V)$ diagram	
has the same probability of being bluer or redder in a different CMD obtained with different data.
To this purpose, we identified the selected stars in {\em u,g} photometry \citep{an08} in Fig.~\ref{CHECK}.
The (a) panel shows a zoom around the RGB,  and again the red line is the fiducial defined as discussed above.
In the following analysis, we considered only those stars in common with the DOLORES photometry and, for the sake
of homogeneity, we kept only stars between 1\arcmin $<R<$4\arcmin~from the cluster center.
That the histogram distributions of the selected RGB stars systematically have red colors demonstrated that we
are seeing a {\em real} feature: no random or systematic errors can explain that the two distribution remain confined in the CMDs 
obtained from independent data sets.

Similar spatial distributions of stars on the bluer  and redder RGBs (panel (a) of Fig.~\ref{CHECK}) 
also indicate that the differential reddening, if any, is not likely the cause of the double RGBs
(see panel (c) in the same figure).
Having demonstrated that the split RGB shown by the $U, (U-V)$ DOLORES photometry is {\em intrinsic},
we named giant stars belonging to the main body of the RGB sequence {\em blue}, while {\em red } are the 
stars located on the anomalous red substructure.
We found that the average color difference for the {\em blue} stars is $\Delta(U-V)_{blue}=-0.005\pm0.016$, significantly different 
from the average color difference for {\em red} stars ($\Delta(U-V)_{red}=-0.251\pm0.017$), which account for only
$\sim$ 4\% of the RGB population in this range of magnitude ($14.5 \leq V \leq 18.5$ mag).
For comparison, $\sim 30\%$ of stars turn out to belong to the blue-RGB in NGC~1851 \citep{lardo12}.
A visual inspection at the CMD of Fig.~3 from \citet{dalessandro09} indeed reinforces our finding and suggests that the 
anomalous RGB is also present in the cluster center. Moreover, \citet{piotto12} claim the presence of a split SGB
for this cluster, with a fainter component remarkably less populous than the brighter one.
We tentatively speculate that, also for M~2, this newly discovered double RGB might be photometrically connected to 
the split SGB, in close analogy to the case of NGC~1851 and M~22. 

\subsection{CH stars along the anomalous RGB}\label{CH}
M~2 contains two CH stars, as discovered by \citet{zinn81} and \citet{smith90}.
These stars show abnormally high CH absorption, together with deep CN bands, compared to other cluster giants.
They are seen in dSph galaxies, and in the Galactic halo, but they
are relatively rare within GCs. 
At present, a handful of stars having enhanced C
and s-process elements have been reported in each of $\omega$~Cen (e.g., \citealp{harding62};
\citealp{bond75}), M~22 \citep{mcclure77}, NGC~1851 \citep{hesser82}, M~55 \citep{smith82}, M~14 \citep{cote97}, and NGC~6426 \citep{sharina12}.
Their spectra usually do not show strong Swan bands of $C_{2}$, the dominate optical spectral features of 
{\em classical} CH stars, suggesting that their anomalous carbon abundances
probably arise through a different mechanism, such as incomplete CN processing \citep{vanture92}.
Indeed, among this sample of
CH-enhanced stars in GCs, only two are likely to be genuine CH stars. Both of these
stars, RGO 55 \citep{harding62} and RGO 70 \citep{dickens72}, are found in $\omega$~Cen.
The surface carbon enhancement of such stars has been attributed to a dredge-up of processed material via mixing or to 
the mass transfer of such material between members of a binary system \citep{mcclure84}.
Moreover, that both $\omega$~Cen and M~22 display heavy element abundance variations 
suggests that in these clusters these CH stars could owe their peculiar chemical pattern to initial enrichment.

\begin{figure}
  \centering
  \resizebox{\hsize}{!}{\includegraphics{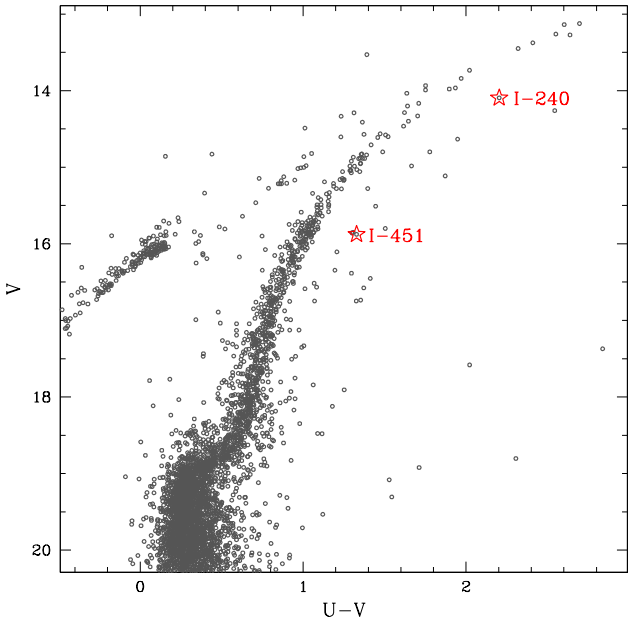}}
\caption{CMD for M2. The location of the carbon stars in
the CMD is indicated by the open stars.}
        \label{CHSTAR}
   \end{figure}

Prompted by these considerations, in Fig.~\ref{CHSTAR} we identified the two CH stars 
discovered by \citet{zinn81} (ID: I-240) and 
\citet{smith90} (ID: I-451) in our $V, U-V$ photometry.
Interestingly enough, both stars belong to the additional RGB, pointing out the anomalous chemical nature
of this redder branch.
Regardless of the exact classification of I-240 and I-451, it is apparent that the anomalous RGB contains a population
of giants that exhibit both a strong CN and strong G band. These stars may be the analogous to 
other CN and CH-strong RGB stars found in $\omega$~Cen, M~22, and NGC~1851 \citep{hesser82}.
Given the peculiarity of other clusters that contain CH stars, it is of extreme interest to investigate the chemical pattern
of stars in this red substructure.
High-resolution spectroscopy of stars in the two distinct groups could be one of the next steps in
deriving the chemical pattern in this cluster, with particular emphasis on the measure of heavy element abundances.

\section{Summary and conclusions}\label{conclusioni}

We have presented low-resolution spectroscopy (R$\simeq$350) of RGB stars in M~2, with the goal
of deriving C abundances (from the G band of CH) and N abundances (from the CN band at $\sim$ 3883\AA).
We were able to measure CH and CN band strengths for 38 giants and derive carbon and nitrogen abundances for 35 stars,
whose spectra were obtained with DOLORES at TNG.
The main results of our analysis can be summarized as follows.

\begin{itemize}
\item We measured the CH and CN band strengths and found large variations ($\simeq$ 0.2-0.3 mag) 
      and a bimodal distribution of CN index strengths (Fig.~\ref{RIDGE}).
      We did report the presence of a clear CH-CN anticorrelation
      over the whole magnitude range (see Fig.~\ref{INDICI}).
\item We used spectral synthesis to measure C and N abundances, and found variations 
      of $\simeq$1 dex and $\simeq$2 dex, respectively, at all luminosities. 
      C and N abundances appear to be anticorrelated, as would be expected from the presence of CN-cycle processing 
      exposed material on the stellar surface (Fig.~\ref{abbondanze}). 
\item Our derived C abundances show a decline with increasing luminosity.
      As far as can be derived from the carbon abundances, the rate of mixing in this cluster is comparable to 
      that of halo field stars and many cluster giants. We found $d[C/Fe]/dM_{V}$=0.21$\pm$0.16, which is very 
      similar within the observational errors to what is found
      among halo field giants and other globulars 
      \citep[e.g., M~3, NGC~6397, and M~13;][]{smith03}.
\item  We distinguished between first and second subpopulations and found that N-poor and N-rich stars 
       are clearly separated into two parallel sequences in the broader giant branch seen in 
       the $V, (U-V)$ diagram, the N-rich stars appearing systematically redder than N-poor ones, a result that is
       strictly analogous to the one of \citet{marino08} for M~4.
\item In addition to these results, we detected an 
      {\em anomalous} substructure beyond the red edge of the
      main body of the RGB (see Fig.~\ref{M2-DOLORES}) from DOLORES $U, V$ photometry.
      When plotting CH stars from the studies of \citet{zinn81} and \citet{smith90} onto the $V, U-V$ DOLORES 
      CMD (see Fig.~\ref{CHSTAR}),
      we found that both stars belong to this additional red RGB.
      These are giants that exhibit both enhanced CH and CN bands, and this evidence perfectly fits the suggestion
      that stars located on the red RGB should have a peculiar chemical nature.
      Moreover, this additional RGB could be connected to the less populated faint SGB detected by \citet{piotto12} in this cluster.
\end {itemize}      

      Among the GCs with photometric evidence of multiple populations, only NGC~1851 and M~22 display a bimodal SGB 
      that is photometrically connected to the split RGB 
      (see \citealp{lardo12} for NGC~1851; \citealp{marino12} for M~22).
      These are rather peculiar clusters  that appear to share
      numerous observational features.
      M~22 shows a spread of about 0.15 dex in metallicity \citep{marino09,alves12}, while
      the presence of an intrinsic iron spread among NGC~1851 stars is still controversial 
      \citep{yong08,carretta10,villanova10}.
      The neutron-capture elements that are mainly produced by
      the $s$-process are also found to have large star-to-star
      abundance variations that are correlated with both [Fe/H] and the abundances
      of light, proton-capture elements \citep{carretta10, marino12}.
      Finally, a spread in the abundances of individual CNO elements
      has been found within both the bright and faint SGB 
      that is also correlated with the variation in
      heavier elements \citep{marino12,lardo12}. 
      Since the Na-O and C-N anticorrelations alone can be considered as signatures of multiple populations
      and both clusters are composed of two different groups of stars with different $s$-element content
      (each of them associated to the bimodal SGB and RGB) possibly with their own Na-O, C-N anticorrelations,
      each group should be the product of multiple star formation episodes.
      Both M~22 and NGC~1851 host not only two subpopulations, but they have experienced a complex
      formation history that resembles the extreme case of $\omega$ Centauri \citep[see][for a discussion]
      {marino12,dacosta11,roederer11,dantona11}.
      The apparent similarity of M~2 to NGC~1851 and M~22 calls for a deeper
      and complete spectroscopic characterization of stars in this poorly studied cluster.

\begin{acknowledgements}
We warmly thank the anonymous referee for a careful reading of the manuscript and
for providing very useful comments.
Support for this work has been provided by the IAC (grant 310394),
and the Education and Science Ministry of Spain (grants AYA2007-3E3506, and AYA2010-16717).
CL acknowledges G. Altavilla for the help in reducing DOLORES data. 
This research has made use of the SIMBAD database, operated at the CDS, Strasbourg, France and of 
the NASA Astrophysical Data System.
This publication makes use of data products from the Two Micron All Sky Survey, 
which is a joint project of the University of Massachusetts and the Infrared Processing and 
Analysis Center/California Institute of Technology, 
funded by the National Aeronautics and Space Administration and the National Science Foundation.

\end{acknowledgements}

\end{document}